\documentclass[onecolumn,showkeys,aps,prd]{revtex4}

\usepackage[utf8]{inputenc}

\usepackage{amsfonts}
\usepackage{latexsym}
\usepackage{amsmath,amssymb,amsthm}
\usepackage{mathtools}
\usepackage{lineno}
\usepackage{bbm}

\renewcommand{\dj}{\mbox{\kern0.6ex\raise0.8ex\hbox{-}\kern-1.4ex d}}
\newcommand{\Dj}{\mbox{\raise0.3ex\hbox{-}\kern-0.4em D}}

\newcommand{\lc}{\varepsilon}

\newcommand{\nablar}{\stackrel{\rightarrow}{\nabla}\!\!{}}
\newcommand{\nablal}{\stackrel{\leftarrow}{\nabla}\!\!{}}
\newcommand{\nablalr}{\stackrel{\leftrightarrow}{\nabla}\!\!{}}
\newcommand{\ds}{\displaystyle}
\newcommand{\del}{\partial}

\newcommand{\diag}{\mathop{\rm diag}\nolimits}

\newcommand{\rmd}{\mathrm{d}}

\newcommand{\realni}{\ensuremath{\mathbb{R}}}

\newcommand{\grasmanovi}{\ensuremath{\mathbb{G}}}

\newcommand{\cF}{{\cal F}}
\newcommand{\cG}{{\cal G}}
\newcommand{\cH}{{\cal H}}

\newcommand{\cL}{{\cal L}}
\newcommand{\cM}{{\cal M}}
\newcommand{\cN}{{\cal N}}

\newcommand{\cR}{{\cal R}}

\newcommand{\g}{\mathfrak{g}}
\newcommand{\h}{\mathfrak{h}}
\renewcommand{\l}{\mathfrak{l}}

\newcommand{\dual}{\text{\boldmath$\,\star$}}

\renewcommand{\leq}{\leqslant}
\renewcommand{\geq}{\geqslant}

\begin{document}

\title{Correspondence between $3BF$ and Einstein-Cartan formulations of quantum gravity}

\author{Pavle Stipsi\'c}
 \email{pstipsic@ipb.ac.rs}
\affiliation{Institute of Physics, University of Belgrade, Pregrevica 118, 11080 Belgrade, Serbia}

\author{Marko Vojinovi\'c}
 \email{vmarko@ipb.ac.rs}
 \thanks{corresponding author}
\affiliation{Institute of Physics, University of Belgrade, Pregrevica 118, 11080 Belgrade, Serbia}


\keywords{quantum gravity, higher gauge theory, $3$-group, $3BF$ action, spin-spin contact interaction, Einstein-Cartan action, path integral quantization}

\begin{abstract}
We construct a correspondence between the quantized constrained $3BF$ theory and the quantized Einstein-Cartan theory with contact spin-spin interaction, both of which describe the Standard Model coupled to Einstein-Cartan gravitational field. First we introduce the expectation values of observables using the path integral formalism for both theories, and then by integrating out some configuration space variables in the quantum $3BF$ theory we obtain the definition of the corresponding observable in the quantum Einstein-Cartan theory with contact interaction. The correspondence is a rather general result, since it can be established without actually performing the detailed quantization of either theory. Finally, we discuss the differences in the predictions of the two theories on the example of the 4-volume density of spacetime, and on the example of gravitational waves.
\end{abstract}

\maketitle

\section{\label{SecI}Introduction}

Quantization of the gravitational field represents one of the main open problems in modern theoretical physics. Over the years, vast research disciplines aiming to formulate a theory of quantum gravity have been proposed and developed, the most prominent ones being string theory \cite{Polchinski1,Polchinski2}, loop quantum gravity \cite{Rovelli2004,Thiemann}, and others. Within the loop quantum gravity framework, one of the promising research directions is based on the idea of covariant quantization, i.e., the quantization by providing a rigorous definition of a path integral for the gravitational field. This is commonly known as the spinfoam quantization programme, and several models of quantized gravitational field have been proposed in the literature \cite{RovelliVidotto2014,MikovicVojinovicBook}.

One of the typical drawbacks of the spinfoam quantization programme is the lack of matter fields in the theory, since the quantization method is adapted to work only for the gravitational field itself. Consequently, various strategies have been proposed to circumvent this issue and generalize the spinfoam quantization programme to include both gravity and matter on equal footing. One such promising generalization has recently been developed, and is based on the application of the so-called higher gauge theory \cite{BaezHuerta,BaezDolan,Baez} and topological quantum field theory techniques \cite{Witten,Atiyah,Quinn,SozerVirelizier,BaezTQFT,Yetter,Porter}. Higher gauge theory is a mathematical framework that provides one with a way to generalize the notion of a gauge symmetry structure by using the higher category theory analogs of Lie groups, called Lie $n$-groups \cite{MikovicVojinovic2012,Li2019,SaemannWolf2014,JurcoEtal2005,HNY2020,HNY2021,SaemannWolf2014b,SWY2021a,SWY2021b,HNY2021b,HNY2021c,JurcoSaemannWolf2016,SaemannWolf2017,JurcoEtal2019b}. In order to successfully implement the spinfoam quantization procedure for a theory that describes both gravity and matter on an equal footing, attention focuses on the notion of a $3$-group, and its corresponding topological action called a $3BF$ action \cite{Radenkovic2019}, which represents a suitable generalization of the well known $BF$ and $2BF$ actions based on an ordinary Lie group and a Lie $2$-group \cite{Plebanski,AKSZ,Pulmann,Cattaneo,baez2000,BFgravity2016,GirelliPfeifferPopescu2008,FariaMartinsMikovic2011,MikovicOliveira2014,Mikovic2015,MOV2016,MOV2019,Asante2020,Girelli2021} (see also \cite{MV2020} for the $4BF$ theory and the $4$-group approach). A number of recent results \cite{MikovicVojinovic2012,Radenkovic2019,Radenkovic2020,Radenkovic2022a,Radenkovic2020proc,Radenkovic2022b,Djordjevic2023,Stipsic2025,Radenkovic2025} have successfully implemented several stages of the spinfoam quantization programme for the constrained $3BF$ theory corresponding to the gravitational field coupled to the full Standard Model. Specifically, general relativity was first rewritten as a constrained $2BF$ model in \cite{MikovicVojinovic2012}, a first result which emphasized the relevance of $nBF$ models for realistic physics. Next, in \cite{Radenkovic2019,Radenkovic2020proc} the theory was extended to a $3BF$ model in order to include matter fields and couple them to gravity, in particular all fields present in the Standard Model (gauge bosons, fermions and scalar fields). After that, the properties of the resulting classical theory have been studied in detail --- the phase space, Hamiltonian analysis, and gauge symmetries of the theory have been discussed in \cite{Radenkovic2020,Radenkovic2022a}, the additional trivial gauge symmetries of the theory were discussed in \cite{Djordjevic2023}, while symmetry breaking and the Higgs mechanism were studied in \cite{Stipsic2025}. Finally, the quantization of the topological $3BF$ theory, and the construction of the topological invariant and its corresponding TQFT, have been done in \cite{Radenkovic2022b,Radenkovic2025}. All this research has demonstrated that the approach to quantum gravity based on the $3$-group and the $3BF$ action is technically tangible --- the goal of providing a rigorous formulation of a theory of quantum gravity with matter seems to be a viable and achievable prospect, using the $3BF$ action and a $3$-group as the starting point.

In this work, we will focus on one interesting property of the $3BF$ formulation of a theory of quantum gravity. Specifically, we will establish a correspondence between the quantization of the suitable constrained $3BF$ action, and the quantization of the standard Einstein-Cartan formulation of general relativity (GR) coupled to the Standard Model. While we will not study the actual details of the quantization of either the $3BF$ theory nor the Einstein-Cartan theory, we will nevertheless be able to introduce a very precise relationship between the two quantum theories. Namely, given any quantum observable that is defined within the context of one specific version of Einstein-Cartan theory (called the Einstein-Cartan contact theory), we will introduce a corresponding observable defined within the context of the quantized $3BF$ theory, such that the expectation values of the two observables match exactly (and vice versa). This correspondence is established at a full nonperturbative level, and it is a surprising feature of the $3BF$ theory that one can in fact formulate such a correspondence using only some general assumptions, i.e., without introducing all details of the actual quantization of either theory.

The obtained correspondence has two important consequences. First, it enables one to in fact {\em define} the quantization of the Einstein-Cartan contact theory coupled to the Standard Model (originally a very hard problem to solve) by passing to the $3BF$ version of the theory, and performing the quantization of the $3BF$ theory instead (a slightly easier problem to solve). In this way, one can circumvent a number of problems that render the quantization of Einstein-Cartan theory non-feasible, and establish it instead by exploiting the obtained correspondence to the quantum $3BF$ theory. The second important consequence of the correspondence is that there exists a regime where the two quantum theories could be experimentally distinguished from each other, at least in principle. To that end, after establishing the correspondence relations as the main result of this work, we will apply those relations to study a few interesting example observables, and discuss in what sense and under which conditions the two quantum theories could be experimentally distinguishable. For example, one can apply the correspondence to compare the magnitude of the strain generated by gravitational waves. In principle, given a source of gravitational waves that is both strong in magnitude and has large quantum uncertainty, one can evaluate the differences in the quantum corrections for the strain amplitude in $3BF$ theory and Einstein-Cartan contact theory, and test them against the experimental data. In this sense, the correspondence predicts observable signatures that distinguish the two theories. Of course, we do not have actual access to a gravitational wave source with the required properties, so any such experimental proposal is still far away from the practical capabilities of current technology, but as a matter of principle, this question can be studied at least theoretically, and it does illustrate the phenomenological significance of the obtained correspondence.

The layout of the paper is as follows. In Section \ref{secII}, we present a short review of the classical $3BF$ and Einstein-Cartan theories, and demonstrate that they give rise to equivalent sets of classical equations of motion. In Section \ref{secIII}, we turn to the main analysis of the expectation values of an arbitrary quantum observable defined in the two quantum theories. After some mathematical preliminaries, we establish a correspondence between the expectation value of the observable in one theory, and the expectation value of a similar observable in the other theory, where ``similar'' means that the observable is weighted by some power of the absolute value of the determinant of the tetrad, $|e|^{\pm M}$. This correspondence is established in a fully nonperturbative way, and represents the main result of the paper. Section \ref{secIV} deals with some illustrative example observables that one can study in order to compare the two quantum theories. First, we discuss the spacetime 4-volume density operator as a simple example, and also the classical limit of the two theories. Then, we discuss the case of the gravitational waves, and give an estimate of how large their quantum uncertainties must be in order to be able to experimentally distinguish between the two quantum theories. In Section \ref{secV} we give our concluding remarks and some topics for future research. The Appendix \ref{AppA} contains proofs of some technical results used in the main text, while the Appendix \ref{AppB} contains some additional mathematical and notational details.

Our notation and conventions are as follows. Spacetime indices, denoted by the mid-alphabet Greek letters $\mu,\nu,\dots$, are raised and lowered by the spacetime metric $g_{\mu\nu}$, once it is defined. The Lorentz metric is denoted as $\eta_{ab} = \diag (-1,+1,+1,+1)$. The indices that are counting the generators of Lie groups $G$, $H$, and $L$ are denoted with initial Greek letters $\alpha, \beta, \dots$, lowercase initial Latin letters $a, b, c,\dots$, and uppercase Latin indices $A,B,C,\dots$, respectively. The generators themselves are typically denoted as $\tau_\alpha$, $t_a$ and $T_A$, respectively. We work in the natural system of units, defined by $c=\hbar=1$ and $G = l_p^2$, where $l_p$ is the Planck length.

The indices which correspond to the Lorentz group are pairs of indices $ab$ and the quantities that depend on them are antisymmetric with respect to their interchange. This means that all independent components of these quantities, according to Einstein summation convention, are summed over twice. Because of this, the result of the summation should be divided by two. Alternatively, in order to avoid this problem, one can introduce the notation $[ab]$ which represents the pair of indices as a single index for which we always assume that $a>b$. Summation over such indices takes into account every independent component precisely once, so it is not necessary to divide the total by two. For example, given some quantity $K^{ab}$, one has
\begin{equation}
K^{[ab]}\sigma_{[ab]}=\frac{1}{2}K^{ab}\sigma_{ab}\,.
\end{equation}
In this work, the square brackets will exclusively denote the pairs of Lorentz indices, rather than the usual antisymmetrization over those indices.

All additional notation and conventions used throughout the paper are explicitly defined in the text where they first appear. See also Appendix \ref{AppB}.

\section{\label{secII}Review of the classical $3BF$ and Einstein-Cartan actions}

In this Section, we will provide a short review of four classical theories that will be relevant for subsequent analysis. We will begin by introducing the topological $3BF$ action, based on the notion of a $3$-group. This will then be employed to introduce the so-called constrained $3BF$ action, which gives rise to physically relevant equations of motion and is one of the main theories that we will subsequently study in Sections \ref{secIII} and \ref{secIV}. Then, we will introduce the standard Einstein-Cartan action, coupled to the Standard Model in the usual way. Finally, we will introduce its corresponding second-order theory, called the Einstein-Cartan contact action. The latter will be the second main theory that we will subsequently study in Sections \ref{secIII} and \ref{secIV}.

\subsection{\label{subsecIIa}Topological $3BF$ action}

In order to introduce the topological $3BF$ action, one first needs to introduce the notion of a strict Lie $3$-group, a generalization of the notion of a Lie group stemming from higher category theory, which is equivalent to the algebraic structure called a Lie 2-crossed module. A Lie 2-crossed module is a triple of Lie groups, $G$, $H$ and $L$, together with two homomorphisms between them,
\begin{equation} \label{eq:jna2}
\partial : H \to G\,, \qquad \delta : L \to H\,,
\end{equation}
the actions of the group $G$ on all three groups,
\begin{equation} \label{eq:jna3}
\triangleright : G \times X \to X\,, \qquad X = G, H, L\,,
\end{equation}
as well as the Peiffer lifting map,
\begin{equation} \label{eq:jna4}
\{ \_ \, , \_\, \}_\mathrm{pf} : H \times H \to L\,.
\end{equation}
All these maps are subject to a certain set of axioms, and together they make up a Lie 2-crossed module, denoted as
\begin{equation} \label{eq:jna5}
(\; L\stackrel{\delta}{\to} H \stackrel{\partial}{\to}G \; , \;\triangleright \;, \; \{\_\,,\_\,\}_\mathrm{pf} \;)\,.
\end{equation}
This structure represents the notion of a $3$-group in the most convenient way for our purposes. An interested reader can find further mathematical details for example in Refs \cite{BaezHuerta,BaezDolan,Radenkovic2019,Radenkovic2022a,Radenkovic2022b,Radenkovic2025,martins2011,SaemannWolf2014b,Wang2014}.

Given the mathematical structure of a $3$-group, it gives rise to a natural choice of an action, called a $3BF$ action (see Appendix \ref{AppB} for a more detailed explanation of the notation used in this section and throughout the text). The $3BF$ action is purely topological, and defined as:
\begin{equation} \label{eq:3BFaction}
S_{3BF}^\text{top}=\int_{\cM_4} \langle B\wedge {\cal F}\rangle_\mathfrak{g}+\langle C\wedge {\cal G}\rangle_\mathfrak{h}+\langle D\wedge {\cal H}\rangle_\mathfrak{l}.
\end{equation} 
The Lagrange multipliers $B$, $C$ and $D$ are two-, one- and zero-forms, and simultaneously they are elements of Lie algebras $\mathfrak{g}$, $\mathfrak{h}$ and $\mathfrak{l}$, corresponding to the Lie groups $G$, $H$ and $L$, respectively. The field strengths $\cF$, $\cG$ and $\cH$ are defined as
\begin{equation} \label{eq:ThreeCurvatureDef}
{\cal F}={\rm d}\alpha+\alpha\wedge \alpha-\partial\beta\,,\qquad {\cal G}={\rm d}\beta+\alpha\wedge^{\triangleright}\beta-\delta\gamma\,,\qquad
{\cal H}={\rm d}\gamma+\alpha\wedge^{\triangleright}\gamma+\{\beta\wedge\beta\}_{\rm pf}\,,
\end{equation}
and they are called fake curvatures for the connection one-form $\alpha$, two-form $\beta$ and three-form $\gamma$, which are also valued in algebras $\mathfrak{g}$, $\mathfrak{h}$ and $\mathfrak{l}$, respectively. Bilinear forms $\langle\_\, ,\_\rangle_{\mathfrak{g}}$, $\langle\_\, ,\_\rangle_{\mathfrak{h}}$  and $\langle\_\, ,\_\rangle_{\mathfrak{l}}$  are assumed to be symmetric, nondegenerate and $G$-invariant, and they map a pair of algebra elements into a real number. Let us also note that, given the structure of the 3-group, one can introduce the notion of a covariant derivative as
\begin{equation} \label{eq:jna8}
\nabla = {\rm d} + \alpha \wedge^{\triangleright}
\end{equation}
in the sense that, when $\nabla$ acts for example on the components $\phi^A$ of the object $\phi \in \l$, the action $\triangleright$ is being applied as the action from the Lie algebra $\g$ to Lie algebra $\l$, giving:
\begin{equation} \label{eq:jna9}
\nabla\phi^A={\rm d}\phi^A+\triangleright_{\alpha B}{}^{A} \,\alpha^{\alpha}\wedge \phi^B\,,
\end{equation}
and similarly for objects that are elements of algebras $\g$ and $\h$. Given this notation, one can rewrite the fake curvatures (\ref{eq:ThreeCurvatureDef}) in terms of ordinary curvatures as:
\begin{equation} \label{eq:ThreeCurvatureRewrite}
{\cal F}=\nabla^2 -\partial\beta\,,\qquad {\cal G}=\nabla\beta -\delta\gamma\,,\qquad
{\cal H}=\nabla\gamma +\{\beta\wedge\beta\}_{\rm pf}\,.
\end{equation}
We point the reader to the Appendix \ref{AppB}, which contains more detailed explanation of the above notation, including some examples.

In order to discuss the field content that corresponds to the Standard Model and Einstein-Cartan gravity, one makes the following choice of the 3-group, called the Standard Model 3-group (see \cite{Radenkovic2019,Radenkovic2020proc,Stipsic2025} for further details). The three Lie groups $G$, $H$ and $L$ are chosen as:
\begin{equation}
G = SO(3,1)\times SU(3)\times SU(2) \times U(1) \,, \qquad H = \mathbb{R}^4\,, \qquad L = \mathbb{C}^4\times\mathbb{G}^{64}\times\mathbb{G}^{64}\times\mathbb{G}^{64}\,.
\end{equation}
The physical interpretation of this choice is as follows. The group $G$ represents the usual Standard Model gauge group, together with the local Lorentz group. The group $H$ represents the spacetime translations, while the group $L$ corresponds to the matter fields. Specifically, $\mathbb{C}^4$ corresponds to the Higgs sector, while the three Grassmann algebras $\mathbb{G}^{64}$ correspond to the three families of fermions.

In order to fully specify the Standard Model $3$-group, one also needs to define all relevant maps. The homomorphisms $\partial$ and $\delta$ are chosen to to be trivial, as well as the Peiffer lifting $\{ \_\, , \_ \,\}_\mathrm{pf}$. Regarding the action $\triangleright$, it is defined as follows. The group $G$ can be naturally split into the Lorentz part $SO(3,1)$ (generators counted using the indices $[ab]$) and the internal gauge part $SU(3)\times SU(2) \times U(1)$ (generators counted collectively using indices $\alpha,\beta,\dots$). The action of $G$ on itself is then given by specifying the action of the Lorentz part on itself and on the internal gauge part, as
\begin{equation} \label{eq:actionOfLorentzPartOfG}
\triangleright_{[ab][cd]}{}^{[ef]}\equiv f_{[ab][cd]}{}^{[ef]}=\frac{1}{2}\left(\eta_{[a|c}\delta_{|b]}^{[f|}\delta_{d}^{|e]}-\eta_{[a|d}\delta_{|b]}^{[f|}\delta_{c}^{|e]}\right)\,, \qquad
\triangleright_{[ab] \beta}{}^{\gamma}=0\,,
\end{equation}
while the action of the internal gauge part on itself and on the Lorentz part is given as
\begin{equation} \label{eq:actionOfInternalPartOfG}
\triangleright_{\alpha \beta}{}^{\gamma} = f_{\alpha\beta}{}^{\gamma} \,, \qquad
\triangleright_{\alpha [ab]}{}^{[cd]}=0\,.
\end{equation}
Further, the action of $G$ on $H$ is specified naturally, assuming that the group $H$ is interpreted as the group of 4-dimensional translations. Then the Lorentz part of $G$ acts in the standard way on translations, while the internal part of $G$ acts trivially:
\begin{equation}\label{eq:actionGOnH}
\triangleright_{[cd]a}{}^{b}=\frac{1}{2}\eta_{a[d|}\delta_{|c]}^b\,, \qquad
\triangleright_{\alpha a}{}^{b}=0\,.
\end{equation}
Finally, the action of the Lorentz and internal subgroups of $G$ on $L$ is also given in a natural way, in accordance with the transformation properties of various fermions and the Higgs scalar. For example, the action of $G$ on left-isospin fermions is given as:
\begin{equation}
\triangleright_{[cd]A}{}^{B}=\left(\sigma_{cd}\right)_A{}^{B}\,, \qquad
\triangleright_{\alpha A}{}^{B}=\frac{1}{2}\left(\sigma_{\alpha}\right)_A{}^{B}\,.
\end{equation}
Here the matrices $\left(\sigma_{\alpha}\right)_A{}^{B}$ are Pauli matrices, and $\left(\sigma_{ab}\right)_A{}^{B}=\frac{1}{4}[\gamma_a,\gamma_b]_A{}^{B}$, where $\gamma_a$ are the standard Dirac matrices satisfying the anticommutation rule $ \gamma_a\gamma_b + \gamma_b\gamma_a = -2 \eta_{ab}$. Here we also introduce $\gamma_5 \equiv - \gamma_0 \gamma_1 \gamma_2 \gamma_3$. In a similar way, one defines the action of group $G$ for all other fermions and scalars in the group $L$, depending on their precise transformation properties (see \cite{Radenkovic2019} for details).

Given the Standard Model $3$-group, one can rewrite the corresponding topological $3BF$ action (\ref{eq:3BFaction}) in the following form:
\begin{equation} \label{eq:3BFforStandardModelPrva}
S_{3BF}^\text{top}=\int B_{\alpha}\wedge F^{\alpha}+ B^{[ab]}\wedge R_{[ab]}+e_a\wedge\nabla\beta^{a}+ \phi^{A}(\nabla\tilde{\gamma})_A+ \bar{\psi}_A(\nablar \gamma)^A-(\bar{\gamma} \nablal)_A\psi^A\,,
\end{equation}
where we have introduced the following new notation. First, $\cF$ is split into the internal symmetry field strength $F^\alpha$ (which is a function of the internal symmetry connection $\alpha^\alpha$) and the Riemann curvature two-form $R_{[ab]}$ (which is a function of the spin connection $\omega^{[ab]}$). The Lagrange multiplier $C$ is rewritten as the tetrad field one-form $e_a$, and the Lagrange multiplier $D$ is rewritten as a tuple of scalar and fermion fields $(\phi^A, \psi^A, \bar{\psi}_A)$. This change of notation also suggests the physical interpretation of the fields in (\ref{eq:3BFaction}).

\subsection{Constrained $3BF$ action}

While the action (\ref{eq:3BFforStandardModelPrva}) does correspond to the Standard Model $3$-group and features all relevant gravitational, gauge and matter fields, it does not provide the correct classical dynamics for those fields. Namely, this action is an example of a {\em topological} $3BF$ action, and as such it has trivial equations of motion, with no propagating degrees of freedom in the theory. In order to remedy this, one introduces additional terms to the action, called {\em simplicity constraint} terms. By conveniently choosing and adding simplicity constraints, we can introduce the full classical action corresponding to the Standard Model coupled to Einstein-Cartan gravity, with the correct classical dynamics. Such an action is then commonly called the {\em constrained} $3BF$ action, and has the following form:
\begin{equation} \label{eq:RealisticAction}
S_{3BF}=S_{3BF}^\text{top}+S_{\text{grav}}+S_{\text{scal}}+S_{\text{Dirac}}+S_{\text{Yang-Mills}}+S_{\text{Higgs}}+S_{\text{Yukawa}}+S_{\text{spin}}+S_{\text{CC}}\,.
\end{equation}
Here we have:
\begin{eqnarray}
S_{3BF}^\text{top}&=&\int B_{\alpha}\wedge F^{\alpha}+ B^{[ab]}\wedge R_{[ab]}+e_a\wedge\nabla\beta^{a}+ \phi^{A}(\nabla\tilde{\gamma})_A+ \bar{\psi}_A(\nablar \gamma)^A-(\bar{\gamma} \nablal)_A\psi^A\,, \label{eq:3BFforStandardModel} \\
S_{\text{grav}}&=&-\int\lambda_{[ab]}\wedge\left(B^{[ab]}-\frac{1}{8\pi l_p^2}\varepsilon^{[ab]cd}e_c\wedge e_d\right)\,, \label{eq:gravConstraint} \\
S_{\text{scal}}&=&\int \tilde{\lambda}^{A}\wedge\left(\tilde{\gamma}_A-H_{abcA}e^a\wedge e^b\wedge e^c\right)+\Lambda^{abA}\wedge\left(H_{abcA}\varepsilon^{cdef}e_d\wedge e_e\wedge e_f-(\nabla\phi)_A\wedge e_a\wedge e_b\right)\,, \label{eq:scalConstraint} \\
S_{\text{Dirac}}&=&\int\bar{\lambda}_A\wedge\left(\gamma^A+\frac{i}{6}\varepsilon_{abcd}e^a\wedge e^b\wedge e^c\left(\gamma^d\psi\right)^{A}\right)-\lambda^{A}\wedge\left(\bar{\gamma}_A-\frac{i}{6}\varepsilon_{abcd}e^a\wedge e^b\wedge e^c \left(\bar{\psi}\gamma^d\right)_A\right)\,, \label{eq:DiracConstraint} \\
S_{\text{Yang-Mills}}&=&\int\lambda^{\alpha}\wedge\left(B_{\alpha}-12{C}_{\alpha\beta}M^{\beta}{}_{ab}e^a\wedge e^b\right)+\zeta_{\alpha}{}^{ab}\left(M^{\alpha}{}_{ab}\varepsilon_{cdef}e^c\wedge e^d \wedge e^e\wedge e^f-F^{\alpha}\wedge e_a\wedge e_b\right)\,, \label{eq:YangMillsConstraintTerm} \\
S_{\text{Higgs}}&=&-\int\frac{2}{4!}\chi\left(\phi^A\phi_A-v^2\right)^2\varepsilon_{abcd}e^a\wedge e^b\wedge e^c\wedge e^d\,, \label{eq:HiggsPotentialConstraint} \\
S_{\text{Yukawa}}&=&-\int \frac{2}{4!}Y_{ABC}\bar{\psi}^A\psi^B\phi^C\varepsilon_{abcd}e^a\wedge e^b\wedge e^c\wedge e^d\,, \label{eq:YukawaConstraint} \\
S_{\text{spin}}&=&\int 2\pi il_p^2\bar{\psi}_A\gamma_5\gamma^a\psi^A\varepsilon_{abcd}e^b\wedge e^c\wedge\beta^d\,, \label{eq:spinConstraint} \\
S_{\text{CC}}&=&-\int\frac{1}{96\pi l_p^2}\Lambda\varepsilon_{abcd}e^a\wedge e^b\wedge e^c\wedge e^d\,. \label{eq:CCconstraint}
\end{eqnarray}
While the form of the full action may appear quite complicated, one can recognize the meaning and purpose of each part of the action, as follows:
\begin{itemize}
\item the topological $3BF$ term (\ref{eq:3BFforStandardModel}) is identical to (\ref{eq:3BFforStandardModelPrva}), tabulating all fields present in the theory (as dictated by the structure of the Standard Model $3$-group),
\item the gravitational constraint term (\ref{eq:gravConstraint}) gives rise to the dynamics of the gravitational degrees of freedom,
\item the scalar constraint (\ref{eq:scalConstraint}) gives rise to the dynamics of massless scalar degrees of freedom,
\item the Dirac constraint (\ref{eq:DiracConstraint}) gives rise to the dynamics of massless fermions,
\item the Yang-Mills constraint (\ref{eq:YangMillsConstraintTerm}) gives rise to the dynamics of massless gauge bosons,
\item the Higgs potential constraint (\ref{eq:HiggsPotentialConstraint}) contains the self-interactions and the mass of the Higgs field,
\item the Yukawa constraint (\ref{eq:YukawaConstraint}) contains the interactions between the Higgs field and fermions, as well as fermion mixing matrices,
\item the spin constraint (\ref{eq:spinConstraint}) is necessary for the appropriate coupling between fermion spins and torsion, and
\item the CC constraint (\ref{eq:CCconstraint}) introduces the cosmological constant.
\end{itemize}
The following free parameters are present in the action:
\begin{itemize}
\item $l_p$ is the Planck length, featuring in $S_\text{grav}$, $S_\text{spin}$ and $S_\text{CC}$,
\item $C_{\alpha\beta}$ represents the gauge coupling constant bilinear form, featuring in $S_\text{Yang-Mills}$,
\item $\chi$ is the coupling constant for the quartic self-interaction of the Higgs field, featuring in $S_\text{Higgs}$,
\item $v$ is the vacuum expectation value of the Higgs field, also featuring in $S_\text{Higgs}$,
\item $Y_{ABC}$ represent the Yukawa couplings and fermion mixing matrices, featuring in $S_\text{Yukawa}$, and
\item $\Lambda$ is the cosmological constant, featuring in $S_\text{CC}$.
\end{itemize}
The topological part $S_{3BF}^\text{top}$ and the constraints $S_\text{scal}$ and $S_\text{Dirac}$ do not contain any free parameters.

Let us discuss the equations of motion (EoMs) for this action. One can obtain the EoMs solved for all Lagrange multiplier fields, in terms of the dynamical fields and their derivatives (see for example \cite{Radenkovic2019,Stipsic2025} for details):
\begin{equation}
\begin{array}{rclcrcl}
    M_{\alpha ab}&=& \displaystyle -\frac{1}{48}\varepsilon_{abcd}F_{\alpha}{}^{\mu\nu}e^{c}{}_{\mu}e^d{}_{\nu}\,,&\hspace*{0.5cm}&
    \zeta^{\alpha ab}&=&\ds \frac{1}{4}{C}_{\beta}{}^{\alpha}\varepsilon^{abcd}F^{\beta}{}_{\mu\nu}e_c{}^{\mu}e_d{}^{\nu}\,, \vphantom{\ds\int} \\
    \lambda_{\alpha \mu\nu}&=&\ds -F_{\alpha \mu\nu}\,,&\hspace*{0.5cm}&
    B_{\alpha \mu\nu}&=&\ds -\frac{e}{2}{C}_{\alpha}{}^{\beta}\varepsilon_{\mu\nu\rho\sigma}F_{\beta}{}^{\rho\sigma}\,, \vphantom{\ds\int} \\
    \lambda_{[ab]\mu\nu}&=&\ds R_{[ab]\mu\nu}\,,&\hspace*{0.5cm}&
    B_{[ab]\mu\nu}&=&\ds \frac{1}{8\pi l_p^2}\varepsilon_{[ab]cd}e^c{}_{\mu}e^d{}_{\nu}\,, \vphantom{\ds\int}  \\
    \tilde{\lambda}^A{}_{\mu}&=&\ds \left(\nabla_{\mu}\phi\right)^A\,,&\hspace*{0.5cm}&
    \tilde{\gamma}^A{}_{\mu\nu\rho}&=&\ds -e\varepsilon_{\mu\nu\rho\sigma}\left(\nabla^{\sigma}\phi\right)^A\,, \vphantom{\ds\int} \\
    H^{abcA}&=&\ds \frac{1}{6e}\varepsilon^{\mu\nu\rho\sigma}\left(\nabla_{\mu}\phi\right)^Ae^a{}_{\nu}e^b{}_{\rho}e^c{}_{\sigma}\,,&\hspace*{0.5cm}&
    \Lambda^{abA}{}_{\mu}&=&\ds \frac{1}{6e}g_{\mu\lambda}\varepsilon^{\lambda \nu\rho\sigma}\left(\nabla_{\nu}\phi\right)^Ae^a{}_{\rho}e^b{}_{\sigma}\,, \vphantom{\ds\int} \\
    \gamma^A{}_{\mu\nu\rho}&=&\ds -i\varepsilon_{abcd}e^a{}_{\mu}e^b{}_{\nu}e^c{}_{\rho}\left(\gamma^d\psi\right)^A\,,&\hspace*{0.5cm}&
    \bar{\gamma}_{A\mu\nu\rho}&=&\ds i\varepsilon_{abcd}e^a{}_{\mu}e^b{}_{\nu}e^c{}_{\rho}\left(\bar{\psi}\gamma^d\right)_A\,, \vphantom{\ds\int} \\
    \lambda^A{}_{\mu}&=&\ds \left(\nablar_{\mu}\psi\right)^A\,,&\hspace*{0.5cm}&
    \bar{\lambda}_{A\mu}&=&\ds \left(\bar{\psi}\nablal_{\mu}\right)_A\,, \vphantom{\ds\int} \\
    \beta^a{}_{\mu\nu}&=&\ds 0\,.  \vphantom{\ds\int} 
    \end{array}
\end{equation}
Next we look at the EoMs for the dynamical fields. The spin connection $\omega^{[ab]}{}_\mu$ is not equivalent to the Levi-Civita connection, since fermionic fields give rise to nonzero torsion. We therefore first split the spin connection into a sum of Ricci rotation coefficients $\Delta^{[ab]}{}_\mu$ and contorsion tensor $K^{[ab]}{}_\mu$:
\begin{equation}  \label{spinskakoneksija}
    \omega^{[ab]}{}_\mu = \Delta^{[ab]}{}_\mu+K^{[ab]}{}_\mu\,.
\end{equation}
Here the Ricci rotation coefficients are given as
\begin{equation}
\Delta^{ab}{}_\mu=\frac{1}{2}\left(c^{abc}-c^{bac}-c^{cab}\right)e_{c\mu}\,,
\end{equation}
where the commutation coefficients are defined as
\begin{equation}
c^{abc}=e^{b\mu} e^{c\nu}\left(\partial_{\mu} e^a{}_\nu-\partial_{\nu} e^a{}_\mu \right)\,.
\end{equation}
The contorsion tensor is given as:
\begin{equation}\label{kontorzija:def}
K^{ab}{}_\mu = \frac{1}{2} \left( T^{cab} + T^{bac} - T^{abc} \right) e_{c\mu} \,.
\end{equation}
Here $T^{abc} \equiv T^a{}_{\mu\nu} e^{b\mu} e^{c\nu}$, where $T^a{}_{\mu\nu}$ are the components of the torsion $2$-form, defined as:
\begin{equation}
T^a \equiv \nabla e^a = \frac{1}{2} T^a{}_{\mu\nu}\, \rmd x^\mu \wedge \rmd x^\nu\,, \qquad T^a{}_{\mu\nu} \equiv \nabla_\mu e^a{}_\nu - \nabla_\nu e^a{}_\mu\,.
\end{equation}
Given all of the above quantities, one can write the EoM for torsion as:
\begin{equation} \label{spinskakoneksija:jna}
    T^a =2\pi i l_p^2\, s^a\,, \qquad s^a \equiv \varepsilon^{abcd} \, e_b\wedge e_c \, \bar{\psi}_A\gamma_5\gamma_d\psi^A\,.
\end{equation}
We can see that the torsion 2-form is proportional to the spin 2-form $s^a$. As we shall see below, this is the same as in Einstein-Cartan gravity. Also, using $(\ref{kontorzija:def})$, the components of the contorsion $1$-form are given as
\begin{equation}
K^{ab}{}_\mu =-2\pi il_p^2\bar{\psi}_A\gamma_5\gamma_d\psi^A\varepsilon^{abcd}e_c{}_{\mu}\,,
\end{equation}
so the relationship between contorsion and torsion simplifies and we obtain:
\begin{equation}
T^a=K^{ab}\wedge e_b\,.
\end{equation}

Next, the Einstein field equation has the usual form:
\begin{eqnarray}\label{ajn}
R_{\mu\nu}-\frac{1}{2}g_{\mu\nu}R+\Lambda g_{\mu\nu}=8\pi l_p^2 \, T_{\mu\nu}\,,
\end{eqnarray}
where the stress-energy tensor is given as:
\begin{eqnarray}
\nonumber
T_{\mu\nu}&=&F^{\alpha}{}_{\mu\rho}C_{\alpha}{}^{\beta}F_{\beta\nu}{}^{\rho}-\frac{1}{4}g_{\mu\nu}F^{\alpha}{}_{\rho\sigma}C_{\alpha}{}^{\beta}F_{\beta}{}^{\rho\sigma}\\
\nonumber
&+&\nabla_{\mu}\phi^A\nabla_{\nu}\phi_A-\frac{1}{2}g_{\mu\nu}\left(\nabla_{\rho}\phi^A\nabla^{\rho}\phi_A+2\chi\left(\phi^A\phi_A-v^2\right)^2\right)\\
&+&\frac{i}{2}\left(\bar{\psi}_A \nablalr_{\mu}\gamma_d\psi^A\right)e{}_{\nu}^d-\frac{1}{2}g_{\mu\nu}\left(i\left(\bar{\psi}_A \nablalr_{\rho}\gamma^d\psi^A\right)e_d{}^{\rho}-2Y_{ABC}\bar{\psi}^A\psi^B\phi^C\right)\,.
\end{eqnarray}
It features three parts, describing the Yang-Mills, scalar and fermion stress-energy, respectively.

EoMs for fermion and scalar fields are
\begin{eqnarray}
\left(i\gamma^{\mu} \nablar_{\mu}\delta_B^A-Y^A{}_{BC}\phi^C\right)\psi^B=0\,,\\
\bar{\psi}_B\left(\delta^B_A i \nablal_{\mu}\gamma^{\mu}+Y_{BAC}\phi^C\right)=0\,,\\
\nabla_{\mu}\nabla^{\mu}\phi^A-4\chi\left(\phi^B\phi_B-v^2\right)\phi^A=0\,,
\end{eqnarray}
while the EoM for Yang-Mills fields is:
\begin{equation}
\nabla_{\mu}F_{\alpha}{}^{\mu\nu}+\frac{1}{2}{C^{-1}}_{\alpha}{}^{\beta}\left(\triangleright_{\beta A B}\left(\phi^A\nabla^{\nu}\phi^B-\phi^B\nabla^{\nu}\phi^A\right)+i\bar{\psi}_A\psi_B\left(\triangleright_{\beta C}{}^A\gamma^{\nu CB}-\gamma^{\nu AC}\triangleright_{\beta C}{}^B\right)\right)=0\,.
\end{equation}
One can observe that all these EoMs correspond precisely to the Standard Model coupled to Einstein-Cartan gravity, along with the cosmological constant term.

From the definition of the action (\ref{eq:RealisticAction}) one can see that the full configuration space of the theory is defined over the non-dynamical Lagrange multiplier fields
\begin{equation}\label{nedinpolja3bf}
    M_{\alpha ab}\,,
    \zeta^{\alpha ab}\,,
    \lambda_{\alpha \mu\nu}\,,
    B_{\alpha \mu\nu}\,,
    \lambda_{[ab]\mu\nu}\,,
    B_{[ab]\mu\nu}\,,
    \tilde{\lambda}^A{}_{\mu}\,,
    \tilde{\gamma}^A{}_{\mu\nu\rho}\,,
    H^{abcA}\,,
    \Lambda^{abA}{}_{\mu}\,,
    \gamma^A{}_{\mu\nu\rho}\,,
    \bar{\gamma}_{A\mu\nu\rho}\,,
    \lambda^A{}_{\mu}\,,
    \bar{\lambda}_{A\mu}\,,
    \beta^a{}_{\mu\nu}\,,
\end{equation}
as well as the dynamical fields
\begin{equation}\label{dinpolja3bf}
\omega^{[ab]}{}_\mu\,, e^a{}_\mu\,, \phi^A\,, \psi^A\,, \bar{\psi}_A\,, \alpha^\alpha{}_\mu \,.
\end{equation}
The distinction between dynamical and non-dynamical fields is a consequence of the equations of motion, since the EoMs for the Lagrange multiplier fields are algebraic equations, while the EoMs for the dynamical fields are partial differential equations. One exception from this convention is the torsion equation (\ref{spinskakoneksija:jna}), which can be explicitly solved for the spin connection as a function of the tetrads and fermion fields, rendering the spin connection as a non-dynamical field as well. This is a well known property in Einstein-Cartan theory, but it is customary to regard the spin connection as a dynamical field nonetheless, because in more general theories in Riemann-Cartan spacetimes it often actually becomes a proper dynamical field \cite{BlagojevicBook,VasilicVojinovic2008}.

\subsection{Einstein-Cartan actions}

The standard Einstein-Cartan (EC) action is typically introduced in the literature (see for example \cite{BlagojevicBook}) as the sum of the actions for the Standard Model minimally coupled to gravity, and the Einstein-Hilbert action for the gravitational field (expressed using the tetrad formalism). In our notation, it reads:
\begin{equation} \label{eq:ECaction}
\begin{array}{lcl}
S_{EC} [e,\omega,\phi, \psi,\bar{\psi},\alpha] & = & \ds \int \frac{1}{16\pi l_p^2} \varepsilon^{abcd}\, R_{ab}\wedge e_c\wedge e_d -F^{\alpha}\wedge C_{\alpha\beta}\dual F^{\beta}-\left(\nabla\phi\right)^A\wedge \left(\dual\nabla\phi\right)_A \vphantom{\ds\int} \\
&& \ds -\frac{i}{6}\varepsilon_{abcd}\,e^a\wedge e^b\wedge e^c\wedge\left[ (\bar{\psi}\nablal)_A\gamma^d\psi^A-\bar{\psi}_A\gamma^{d} (\nablar\psi )^A\right] \vphantom{\ds\int} \\
&& \ds -\frac{1}{12}\vphantom{\int}\left[\chi\left(\phi^A\phi_A-v^2\right)^2+Y_{ABC}\bar{\psi}^A\psi^B\phi^C+\frac{\Lambda}{8\pi l_p^2}\right]\varepsilon_{abcd} \, e^a\wedge e^b\wedge e^c\wedge e^d\,. \vphantom{\ds\int} \\
\end{array}
\end{equation}
Here $\dual F$ denotes the Hodge dual of the $2$-form $F$, and similarly for the $1$-form $\nabla\phi$:
\begin{equation} \label{eq:HodgeDualDef}
  \dual F^{\alpha}=\frac{1}{4}F^{\alpha}{}_{cd}\varepsilon^{abcd}e_a\wedge e_b\,, \qquad
  \left(\dual\nabla\phi\right)_A=\frac{1}{3!}\left(\nabla_d\phi\right)_A\varepsilon^{dabc}e_a\wedge e_b\wedge e_c\,.
\end{equation}
The configuration space of this theory is equivalent to the configuration space of the dynamical fields of the $3BF$ theory $(\ref{dinpolja3bf})$, where the spin connection again satisfies the algebraic equation
\begin{equation} \label{eq:SpinConnectionRicciCoef}
\omega_{ab\mu}=\Delta_{ab\mu}-2\pi il_p^2\bar{\psi}_A\gamma_5\gamma^d\psi^A\varepsilon_{abcd}e^c{}_{\mu}\,,
\end{equation}
obtained from the equation of motion for torsion analogous to equation $(\ref{spinskakoneksija:jna})$, so it does not represent a dynamical field. Because of this, by substituting this algebraic relation back into the action, it is possible to construct an equivalent classical theory with the reduced configuration space and equivalent equations of motion. Such equivalent theory is usually called the {\em second order theory} in the literature \cite{BlagojevicBook}. This substitution is performed by explicitly partitioning the spin connection into the contorsion and Ricci coefficients in the action, and by separating the contributions of the individual terms. This operation is equivalent to the following substitution:
\begin{equation}
\nabla_{(\omega)}=\nabla_{(\Delta)}+\frac{1}{2}K^{ab}\sigma_{ab}
\end{equation}
In the action of the EC theory there are only two terms which depend on the spin connection, namely the term with the Riemann tensor and the Dirac Lagrangian term. After the substitution, these terms become:
\begin{eqnarray}
\nonumber
\vphantom{\int}R_{ab}&=&{\rm d}\omega_{ab}+\left(\omega\wedge\omega\right)_{ab}={\rm d}\Delta_{ab}+\left(\Delta\wedge\Delta\right)_{ab}+{\rm d}K_{ab}+\left(\Delta\wedge K\right)_{ab}+\left(K\wedge \Delta\right)_{ab}+K_a{}^{c}\wedge K_{cb}\\
\vphantom{\int}&=&R_{ab}(\Delta)+\left(\nabla_{(\Delta)}K\right)_{ab}+K_a{}^{c}\wedge K_{cb}\,,
\end{eqnarray}
and
\begin{eqnarray}
\nonumber
\left(\bar{\psi}\overleftarrow{\nabla}\right)_A\gamma^d\psi^A-\bar{\psi}_A\gamma^d\left(\nabla\psi\right)^A=\left(\bar{\psi}\overleftarrow{\nabla}_{(\Delta)}\right)_A\gamma^d\psi^A-\bar{\psi}_A\gamma^d\left(\nabla_{(\Delta)}\psi\right)^A-\frac{1}{2}\bar{\psi}_A\left\{\sigma^{ab}\,,\gamma^d\right\}\psi^A K_{ab}\\
=\left(\bar{\psi}\overleftarrow{\nabla}_{(\Delta)}\right)_A\gamma^d\psi^A-\bar{\psi}_A\gamma^d\left(\nabla_{(\Delta)}\psi\right)^A+\frac{1}{2}\varepsilon^{abcd}\bar{\psi}_A\gamma_{5}\gamma_c\psi^A K_{ab}\,.
\end{eqnarray}
When these terms are substituted back into the standard EC action, after some algebra one obtains the action of the same form as the initial one, and with an extra term representing the contact spin-spin interaction and the fixed spin connection $\Delta_{ab}$. This extra term, expressed using contorsion, is $\frac{1}{8\pi l_p^2}K^{ab}\wedge \dual K_{ab}$, so using the relation between the spin tensor and contorsion from the standard EC theory one can eliminate the contorsion from the action:
\begin{equation} \label{eq:ECCaction}
\begin{array}{lcl}
S_{ECC} [e,\phi, \psi,\bar{\psi},\alpha] & = & \ds \int \frac{1}{16\pi l_p^2} \varepsilon^{abcd}\, R_{ab}\wedge e_c\wedge e_d -F^{\alpha}\wedge C_{\alpha\beta}\dual F^{\beta}-\left(\nabla\phi\right)^A\wedge \left(\dual\nabla\phi\right)_A \vphantom{\ds\int} \\
&& \ds -\frac{i}{6}\varepsilon_{abcd}\,e^a\wedge e^b\wedge e^c\wedge\left[ (\bar{\psi}\nablal)_A\gamma^d\psi^A-\bar{\psi}_A\gamma^{d} (\nablar\psi )^A\right] \vphantom{\ds\int} +\frac{3}{\left(4\pi l_p^2\right)^3}s^{a}\wedge \dual s_{a}\\
&& \ds -\frac{1}{12}\vphantom{\int}\left[\chi\left(\phi^A\phi_A-v^2\right)^2+Y_{ABC}\bar{\psi}^A\psi^B\phi^C+\frac{\Lambda}{8\pi l_p^2}\right]\varepsilon_{abcd} \, e^a\wedge e^b\wedge e^c\wedge e^d\,. \vphantom{\ds\int} \\
\end{array}
\end{equation}
In this fashion, one obtains the Einstein-Cartan contact action (ECC), with a fourth degree contact interaction terms between fermions, featuring within the term $s^{a}\wedge \dual s_{a}$. The equations of motion of this theory are equivalent to the equations of motion for the standard EC theory and the constrained $3BF$ theory, but the configuration space is further reduced and equal to the dynamical configuration space:
\begin{equation}\label{dinpoljaecc}
e^a{}_\mu\,, \phi^A\,, \psi^A\,, \bar{\psi}_A\,, \alpha^\alpha{}_\mu \,.
\end{equation}

This concludes the review of the topological and constrained $3BF$ actions, as well as the EC and ECC actions, regarded as classical theories. In what follows, we will focus our attention to the constrained $3BF$ action (from now on just called $3BF$ action for simplicity), and the ECC action. Specifically, in the next Section we will explore the relationship between their corresponding quantum theories, namely the quantized $3BF$ theory and the quantized ECC theory. This will be done within in a framework which consistently defines the quantization of both theories in the same way, so that they can be compared.

\section{\label{secIII}Quantum observables}

Although the $3BF$ and ECC theories are classically equivalent, their quantization may give rise to potentially different quantum theories. In this Section we will establish a correspondence between the expectation values of observables in the two quantum theories, demonstrating that although the theories are not fully equivalent at the quantum level, there exists a well-defined correspondence between them.

The process of construction of quantum theories requires the generalization of mathematical results related to multiple integrals over real and Grassmann numbers to the corresponding functional integrals over bosonic and fermionic fields. Therefore, in the first Subsection we will provide a review of the integrals that will be generalized to functional level. In the second Subsection we will apply those integrals to the definition of the expectation value of an arbitrary observable in the quantum $3BF$ theory, step by step, so that the obtained result can be interpreted as the expectation value of a related observable in the quantum ECC theory. In this way, we will construct a correspondence between the two quantum theories, at the full nonperturbative level.

\subsection{Mathematical preliminaries}

Let us begin from the definitions of certain mathematical identities. We will discuss a total of four generalizations of the properties of the Dirac delta function in several special cases, as well as one identity related to the Stokes theorem. The identities can be classified into two groups, bosonic and fermionic. The two bosonic identities can be obtained by generalizing the following properties of the Dirac delta function to the bosonic fields. The first identity is
\begin{equation}
\int_\realni dy \, e^{iyF} = 2\pi\, \delta(F)\,, \qquad F\in\realni\,,
\end{equation}
based on which one can obtain the following multiple integral:
\begin{equation}
\int_\realni dy\int_{\realni^n} dx_k\, H(x_k)\,e^{i\left(yF\left(x_k\right)+G(x_k)\right)}=2\pi\int_{\realni^n} dx_k \,H(x_k)\delta\left(F(x_k)\right)\,e^{iG(x_k)}\,.
\end{equation}
Then, generalizing this identity to the functional level, one obtains:
\begin{equation}\label{bozonska_delta_def}
\int D\varphi D\phi_k \,H(\phi_k)\,e^{i\int\left(\varphi \wedge F(\phi_k)+G(\phi_k)\right)}={\cal N}\int D\phi_k \,H(\phi_k)\delta\left(F(\phi_k)\right)\,e^{i\int G(\phi_k)}\,.
\end{equation}
The second identity that we need is:
\begin{equation}
\int_\realni dy \, \delta(yF - G) \, H(y) = \frac{H(G/F)}{|F|}\,, \qquad F,G\in \realni\,,
\end{equation}
which can be easily proved using a simple change of variables. The corresponding multiple integral is
\begin{equation}
\int_\realni dy\int_{\realni^n}dx_k\,\delta\left(y^{aB}F_{B}{}^{A}(x_k)-G^{aA}(x_k)\right)H(y,x_k)=\int_{\realni^n} dx_k\frac{H\left(G^{aB}(x_k)F^{-1}{}_B{}^A(x_k),x_k\right)}{\left|F(x_k)\right|^{|a|}}\,,
\end{equation}
while the generalization to the functional level is given as:
\begin{equation}\label{bozonska_delta_int}
\int D\varphi D\phi_k\,\delta\left(\varphi^{aB}F_{B}{}^{A}(\phi_k)-G^{aA}(\phi_k)\right)H(\varphi,\phi_k)=\int D\phi_k \frac{1}{\left|F(\phi_k)\right|^{|a|}}H\left(G^{aB}(\phi_k)F^{-1}{}_B{}^A(\phi_k),\phi_k\right)\,,
\end{equation}
where $F_B{}^{A}(\phi_k)$ is an arbitrary invertible matrix function, $G^{aA}(\phi_k)$ and $H(\varphi,\phi_k)$ are arbitrary functions, and $|a|$ denotes the number of possible values of the index $a$.

Next, we need an identity related to the Stokes theorem, namely
\begin{equation}\label{stoksovopravilo}
\begin{array}{l}
\ds \int D\varphi D\phi_k \,H(\varphi,\phi_k)\,e^{i\int\left(\nabla \varphi\right)\wedge E(\phi_k)+\varphi\wedge F(\phi_k)+G(\phi_k)}\,\\
\ds \hphantom{mmmmm} =\int D\varphi_\partial D\phi_{k\partial}\,e^{i\oint \varphi_{\partial}\wedge E(\phi_{k\partial})}\int D\varphi D\phi_k \,H(\varphi,\phi_k)\,e^{i\int (-1)^{p-1}\varphi\wedge \nabla E(\phi_k)+ \varphi\wedge F(\phi_k)+G(\phi_k)}\,, \\
\end{array}
\end{equation}
which holds equally for both bosonic and fermionic fields. Here it is assumed that $\varphi$ is a $p$-form, while $\varphi_\partial$ and $\phi_{k\partial}$ are values of the fields on the integration boundary. Note that the sole purpose of this identity is to move the covariant derivative $\nabla$ from acting on $\varphi$ to acting on $E(\phi_k)$.

Finally, the two fermionic identities which can be equivalently generalized from Grassmann numbers to Grassmann fields are given as
\begin{equation}\label{fermionska_delta_def}
\begin{array}{l}
\ds\int_{\grasmanovi^n} d\theta_1 d\theta_2 \dots d\theta_n \; e^{i\theta_1\left(\theta_2-\theta_3-\dots-\theta_k\right)}\;F(\theta_2,\dots,\theta_n)\\
\ds \hphantom{mmmmm} =(-1)^{n-1} i\int_{\grasmanovi^{n-1}} d\theta_2\dots d\theta_n\;\delta(\theta_2-\theta_3-\dots -\theta_k)\;F(\theta_2,\dots,\theta_n)\,, \\
\end{array}
\end{equation}
and
\begin{equation}\label{fermionska_delta_int}
\begin{array}{l}
\ds \int_{\realni^m}d^m y \int_{\realni^m}d^m x\int_{\grasmanovi^k} d\theta_1 d\theta_2\dots d\theta_k  \int_{\grasmanovi^{n-k}} d\theta_{k+1} \dots d\theta_n\; e^{iy_a\left(x^a-M^{aij}\theta_{i}\theta_j\right)}\; F(x,\theta_1,\dots,\theta_n)\\
\ds \hphantom{mmmmm} =(2\pi)^m \int_{\realni^m}d^m x\int_{\grasmanovi^k} d\theta_1 d\theta_2 \dots d\theta_k\int_{\grasmanovi^{n-k}} d\theta_{k+1}\dots d\theta_n\; \prod_{a=1}^m \delta\left(x^a-M^{aij}\theta_{i}\theta_{j}\right)\; F(x,\theta_1,\dots,\theta_n)\,. \\
\end{array}
\end{equation}
For proof see Appendix \ref{AppA}. The corresponding functional identities for the Grassmann fields look the same as the above two identities, up to the replacement of the integration measure $d\theta \to D\theta$, the overall coefficient $(2\pi)^m \to \cN$, and the notation $\prod \delta(x^a) \to \delta(\phi)$. One should note that the last identity contains the Dirac delta function with a combination of real and Grassmann numbers where the order of integration is important, namely, one must first integrate over the real numbers $x^a$, and only afterwards over the Grassmann numbers. Also, the Dirac delta function for the Grassmann numbers is odd (i.e., skew-Hermitian) because of the anticommutativity of the Grassmann numbers, so it always appears paired with an imaginary unit $i \equiv \sqrt{-1}$ in equation $(\ref{fermionska_delta_def})$.

\subsection{Expectation values of the observables}

The correspondence between two quantum theories can be obtained by comparing the expectation values of the observables between the two theories. To that end, one defines the expectation values as
\begin{equation}\label{definicijaopservabli}
\langle F\rangle_{3BF}=\frac{1}{Z_{3BF}}\int D\phi_i \; F(\phi_k) \;e^{iS_{3BF}[\phi_i]}\,,\qquad\langle F\rangle_{ECC}=\frac{1}{Z_{ECC}}\int D\phi_i \; F(\phi_k)\; e^{iS_{ECC}[\phi_i]}\,,
\end{equation} 
where the state sums are given as
\begin{equation}
Z_{3BF}=\int D\phi_i \; e^{iS_{3BF}[\phi_i]}\,,\qquad Z_{ECC}=\int D\phi_i \; e^{iS_{ECC}[\phi_i]}\,.
\end{equation}
Here at the beginning it is important to make two comments. First, for the purpose of subsequent analysis one does not need to discuss the precise definition of the path integral itself, which means that we do not need to specify the quantum $3BF$ and ECC theories explicitly. The only requirement that we assume is that the measures in the path integrals are defined in the same way in both theories, and that they are defined such that the functional identities (\ref{bozonska_delta_def}), (\ref{bozonska_delta_int}), (\ref{stoksovopravilo}), (\ref{fermionska_delta_def}) and (\ref{fermionska_delta_int}) hold. In this way, we can discuss the properties and compare the expectation values of the observables in two quantum theories in the full nonperturbative regime, despite the fact that we do not specify the details of the quantizations of the two theories.

Second, the fields $\phi_i$ for $3BF$ theory and ECC theory belong to their corresponding configurations spaces (\ref{nedinpolja3bf})-(\ref{dinpolja3bf}) and (\ref{dinpoljaecc}), respectively. It should be clear that the observables $F(\phi_k)$ in (\ref{definicijaopservabli}) can be compared only if they both live in the common configuration subspace of both theories, or in other words, if the observable $F$ depends only on the fields $\phi_k$ belonging to the reduced configuration subspace (\ref{dinpoljaecc}) which is defined for the ECC theory.

In what follows, we will apply the functional identities (\ref{bozonska_delta_def}), (\ref{bozonska_delta_int}), (\ref{stoksovopravilo}), (\ref{fermionska_delta_def}) and (\ref{fermionska_delta_int}), step by step, in order to reduce the definition of the expectation value of the observable $F$ in quantum $3BF$ theory, to the definition of the expectation value of the corresponding observable in quantum EC theory, and then in the quantum ECC theory.

We begin by explicitly writing the product of the state sum and the expectation value of the observable $F$ in quantum $3BF$ theory. According to the definition $(\ref{definicijaopservabli})$, it is equal to:
\begin{eqnarray} \label{eq:initialstep}
\nonumber
 Z_{3BF} \langle F\rangle_{3BF} & = & {\cal N}\int D\alpha D\omega D\beta De D\tilde{\gamma} D\gamma D\bar{\gamma} D\phi D\psi D\bar{\psi} DB D\tilde{\lambda} D\lambda D\bar{\lambda} D\Lambda D\zeta DH DM\\
\nonumber
&&\mathrm{exp}\left(i\int B_{\alpha}\wedge F^{\alpha}+B_{[ab]}\wedge R^{[ab]}+e_a\wedge\left(\nabla\beta\right)^{a}+\phi^A\left(\nabla\tilde{\gamma}\right)_A+\bar{\psi}_A\left(\nabla\gamma\right)^A-\left(\bar{\gamma}\overleftarrow{\nabla}\right)_A\psi^A\right.\\
\nonumber
&&+\lambda^{\alpha}\wedge\left(B_{\alpha}-12C_{\alpha\beta}M^{\beta}{}_{ab}e^a\wedge e^b\right)-\lambda_{[ab]}\wedge\left(B^{[ab]}-\frac{1}{8\pi l_p^2}\varepsilon^{[ab]cd}e_c\wedge e_d\right)\\
\nonumber
&&+\tilde{\lambda}^A\wedge\left(\tilde{\gamma}_A-H_{abcA}e^a\wedge e^b\wedge e^c\right)+\bar{\lambda}_A\wedge\left(\gamma^A+\frac{i}{6}\varepsilon_{abcd}e^a\wedge e^b\wedge e^c\left(\gamma^d\psi\right)^A\right)\\
\nonumber
&&-\lambda^A\wedge\left(\bar{\gamma}_A-\frac{i}{6}\varepsilon_{abcd}e^a\wedge e^b\wedge e^c\left(\bar{\psi}\gamma^d\right)_A\right)+2\pi i l_p^2 \bar{\psi}_A\gamma_5\gamma^a\psi^A
\varepsilon_{abcd}e^b\wedge e^c\wedge \beta^d\\
\nonumber
&&+\zeta_{\alpha}{}^{ab}\left(M^{\alpha}{}_{ab}\varepsilon_{cdef}e^c\wedge e^d\wedge e^e\wedge e^f-F^{\alpha}\wedge e_a\wedge e_b\right)\vphantom{\int}\\
\nonumber
&&+\Lambda^{abA}\wedge\left(H_{abcA}\varepsilon^{cdef}e_d\wedge e_e\wedge e_f-\left(\nabla\phi\right)_A\wedge e_a\wedge e_b\right)\vphantom{\int}\\
\nonumber
&&-\frac{1}{12}\left.\vphantom{\int}\left(\chi\left(\phi^A\phi_A-v^2\right)^2+Y_{ABC}\bar{\psi}^A\psi^B\phi^C+\frac{\Lambda}{8\pi l_p^2}\right)\varepsilon_{abcd}e^a\wedge e^b\wedge e^c\wedge e^d\right)\\
&&F\left(\phi_k\right)\,.
\end{eqnarray}
Applying the Stokes theorem functional identity ($\ref{stoksovopravilo}$) over the terms in the second row, we remove the covariant derivative from the variables $\beta_a$, $\bar{\gamma}_A$, $\gamma^A$ and $\tilde{\gamma}^A$, in order to be able to perform the integration over these variables in the next step. We obtain:
\begin{eqnarray} \label{eq:stepone}
\nonumber
 Z_{3BF} \langle F \rangle_{3BF} & = & {\cal N}\int D\beta_\partial De_\partial D\tilde{\gamma}_\partial D\gamma_{\partial} D\bar{\gamma}_\partial D\phi_\partial D\psi_\partial D\bar{\psi}_\partial\,e^{i\oint\phi^A_\partial\tilde{\gamma}_{A\partial}+\bar{\psi}_{A\partial}\gamma^A_\partial+\bar{\gamma}_{A\partial}\psi^A_\partial-e_{a\partial}\wedge\beta^a_\partial}\vphantom{\int}\\
\nonumber
&&\int D\alpha D\omega D\beta De D\tilde{\gamma} D\gamma D\bar{\gamma} D\phi D\psi D\bar{\psi} DB D\tilde{\lambda} D\lambda D\bar{\lambda} D\Lambda D\zeta DH DM \\
\nonumber
&&\mathrm{exp}\left(i\int B_{\alpha}\wedge F^{\alpha}+B_{[ab]}\wedge R^{[ab]}+\left(\nabla e\right)_a\wedge\beta^{a}-\left(\nabla\phi\right)^A\wedge\tilde{\gamma}_A-\left(\bar{\psi}\overleftarrow{\nabla}\right)_A\wedge\gamma^A+\bar{\gamma}_A\wedge\left(\nabla\psi\right)^A\right.\\
\nonumber
&&+\lambda^{\alpha}\wedge\left(B_{\alpha}-12C_{\alpha\beta}M^{\beta}{}_{ab}e^a\wedge e^b\right)-\lambda_{[ab]}\wedge\left(B^{[ab]}-\frac{1}{8\pi l_p^2}\varepsilon^{[ab]cd}e_c\wedge e_d\right)\\
\nonumber
&&+\tilde{\lambda}^A\wedge\left(\tilde{\gamma}_A-H_{abcA}e^a\wedge e^b\wedge e^c\right)+\bar{\lambda}_A\wedge\left(\gamma^A+\frac{i}{6}\varepsilon_{abcd}e^a\wedge e^b\wedge e^c\left(\gamma^d\psi\right)^A\right)\\
\nonumber
&&-\lambda^A\wedge\left(\bar{\gamma}_A-\frac{i}{6}\varepsilon_{abcd}e^a\wedge e^b\wedge e^c\left(\bar{\psi}\gamma^d\right)_A\right)+2\pi i l_p^2 \bar{\psi}_A\gamma_5\gamma^a\psi^A
\varepsilon_{abcd}e^b\wedge e^c\wedge \beta^d\\
\nonumber
&&+\zeta_{\alpha}{}^{ab}\left(M^{\alpha}{}_{ab}\varepsilon_{cdef}e^c\wedge e^d\wedge e^e\wedge e^f-F^{\alpha}\wedge e_a\wedge e_b\right)\vphantom{\int}\\
\nonumber
&&+\Lambda^{abA}\wedge\left(H_{abcA}\varepsilon^{cdef}e_d\wedge e_e\wedge e_f-\left(\nabla\phi\right)_A\wedge e_a\wedge e_b\right)\vphantom{\int}\\
\nonumber
&&-\frac{1}{12}\left.\vphantom{\int}\left(\chi\left(\phi^A\phi_A-v^2\right)^2+Y_{ABC}\bar{\psi}^A\psi^B\phi^C+\frac{\Lambda}{8\pi l_p^2}\right)\varepsilon_{abcd}e^a\wedge e^b\wedge e^c\wedge e^d\right)\\
&&F(\phi_k)\,.
\end{eqnarray}
By careful inspection of the above equation one can note that, in the case when the observable $F(\phi_k)$ does not depend on the fields at the boundary, the integration over the boundary gives no contribution at all. Also, formulated in this way, the $3BF$ theory enforces the restrictions that the matter fields and tetrad fields must be zero on the manifold boundary. These restrictions can be removed by adding appropriate boundary terms to the classical $3BF$ action. The detailed discussion of the boundary conditions is given in Subsection \ref{SubSecBoundary} below.

Next, we perform the integration over the fields $\beta_a$, $B_{[ab]}$, $B_{\alpha}$, $\tilde{\gamma}^A$, $\Bar{\gamma}_A$, $\gamma_A$, $\zeta_{\alpha}{}^{ab}$ and $\Lambda^{abA}$ by applying the functional identities ($\ref{bozonska_delta_def}$) and $(\ref{fermionska_delta_def})$, which gives:
\begin{eqnarray} \label{eq:steptwo}
\nonumber
 Z_{3BF} \langle F \rangle_{3BF} & = & {\cal N}\int De_\partial D\phi_\partial D\psi_\partial D\bar{\psi}_\partial\,\delta(\phi_{\partial})\delta(\psi_\partial)\delta(\bar{\psi}_\partial)\delta(e_\partial)\\
\nonumber
&&\vphantom{\int}\int D\alpha D\omega De D\phi D\psi D\bar{\psi} D\tilde{\lambda} D\lambda D\bar{\lambda} DH DM\\
\nonumber
&&\delta\left(F^{\alpha}+\lambda^{\alpha}\right)\delta\left(R_{[ab]}-\lambda_{[ab]}\right)\delta\left(\tilde{\lambda}_A-\left(\nabla\phi\right)_A\right)\delta\left(\bar{\lambda}_A-\left(\bar{\psi}\overleftarrow{\nabla}\right)_A\right)\delta\left(\lambda^A-\left(\nabla\psi\right)^A\right)\\
\nonumber
&&\mathrm{exp}\left(i\int-12\lambda^{\alpha}\wedge C_{\alpha\beta}M^{\beta}{}_{ab}e^a\wedge e^b+\lambda_{[ab]}\wedge\frac{1}{8\pi l_p^2}\varepsilon^{[ab]cd}e_c\wedge e_d-\tilde{\lambda}^A\wedge H_{abcA}e^a\wedge e^b\wedge e^c\right.\\
\nonumber
&&+\bar{\lambda}_A\wedge\frac{i}{6}\varepsilon_{abcd}e^a\wedge e^b\wedge e^c\left(\gamma^d\psi\right)^A+\lambda^A\wedge\frac{i}{6}\varepsilon_{abcd}e^a\wedge e^b\wedge e^c\left(\bar{\psi}\gamma^d\right)_A\\
\nonumber
&&-\frac{1}{12}\left.\vphantom{\int}\left(\chi\left(\phi^A\phi_A-v^2\right)^2+Y_{ABC}\bar{\psi}^A\psi^B\phi^C+\frac{\Lambda}{8\pi l_p^2}\right)\varepsilon_{abcd}e^a\wedge e^b\wedge e^c\wedge e^d\right)\\
\nonumber
&&\delta\left(M^{\alpha}{}_{ab}\varepsilon_{cdef}e^c\wedge e^d\wedge e^e\wedge e^f-F^{\alpha}\wedge e_a\wedge e_b\right)\vphantom{\int}\\
\nonumber
&&\delta\left(H_{abcA}\varepsilon^{cdef}e_d\wedge e_e\wedge e_f-\left(\nabla\phi\right)_A\wedge e_a\wedge e_b\right)\\
\nonumber
&&\delta\left(\left(\nabla e\right)_a-2\pi il_p^2\bar{\psi}_A\gamma_5\gamma^d\psi^A\varepsilon_{abcd}e^b\wedge e^c\right)\vphantom{\int}\\
&&F(\phi_k)\,.
\end{eqnarray}
Then, further integration over $\lambda^{\alpha}$, $\lambda_{[ab]}$, $\tilde{\lambda}_A$, $\bar{\lambda}_A$ and $\lambda^A$ removes the Dirac delta terms from the third row, thus giving the following:
\begin{eqnarray} \label{eq:stepthree}
\nonumber
 Z_{3BF} \langle F \rangle_{3BF} & = & {\cal N}\int De_\partial D\phi_\partial D\psi_\partial D\bar{\psi}_\partial\,\delta(\phi_{\partial})\delta(\psi_\partial)\delta(\bar{\psi}_\partial)\delta(e_\partial)\\
\nonumber
&&\vphantom{\int}\int D\alpha D\omega De D\phi D\psi D\bar{\psi} DH DM\\
\nonumber
&&\mathrm{exp}\left(i\int 12F^{\alpha}\wedge C_{\alpha\beta}M^{\beta}{}_{ab}e^a\wedge e^b+R_{ab}\wedge\frac{1}{16\pi l_p^2}\varepsilon^{abcd}e_c\wedge e_d-\left(\nabla\phi\right)^A\wedge H_{abcA}e^a\wedge e^b\wedge e^c\right.\\
\nonumber
&&-\frac{i}{6}\varepsilon_{abcd}e^a\wedge e^b\wedge e^c\wedge\left(\left(\bar{\psi}\overleftarrow{\nabla}\right)_A\gamma^d\psi^A-\bar{\psi}_A\gamma^{d}\left(\nabla\psi\right)^A\right)\\
\nonumber
&&-\frac{1}{12}\left.\vphantom{\int}\left(\chi\left(\phi^A\phi_A-v^2\right)^2+Y_{ABC}\bar{\psi}^A\psi^B\phi^C+\frac{\Lambda}{8\pi l_p^2}\right)\varepsilon_{abcd}e^a\wedge e^b\wedge e^c\wedge e^d\right)\\
\nonumber
&&\delta\left(M^{\alpha}{}_{ab}\varepsilon_{cdef}e^c\wedge e^d\wedge e^e\wedge e^f-F^{\alpha}\wedge e_a\wedge e_b\right)\vphantom{\int}\\
\nonumber
&&\delta\left(H_{abcA}\varepsilon^{cdef}e_d\wedge e_e\wedge e_f-\left(\nabla\phi\right)_A\wedge e_a\wedge e_b\right)\\
\nonumber
&&\delta\left(2\pi il_p^2\bar{\psi}_A\gamma_5\gamma^d\psi^A\varepsilon_{abcd}e^b\wedge e^c-\left(\nabla e\right)_a\right)\vphantom{\int}\\
&&F(\phi_k)\,.
\end{eqnarray}
The multipliers $M^{\alpha}{}_{ab}$ and $H_{abcA}$, which have so far not been integrated over, are related to the Hodge duals of the field strengths $F^{\alpha}$ and $\left(\nabla\phi\right)_A$, respectively. Applying the functional identity ($\ref{bozonska_delta_int}$) one can integrate out also these remaining multipliers, in favor of the Hodge duals (\ref{eq:HodgeDualDef}), giving:
\begin{eqnarray} \label{eq:stepfour}
\nonumber
 Z_{3BF} \langle F \rangle_{3BF} & = & {\cal N}\int De_\partial D\phi_\partial D\psi_\partial D\bar{\psi}_\partial\,\delta(\phi_{\partial})\delta(\psi_\partial)\delta(\bar{\psi}_\partial)\delta(e_\partial)\vphantom{\int} \\
\nonumber
&& \int D\alpha D\omega De D\phi D\psi D\bar{\psi}\frac{1}{|e|^{(|\alpha|+|A|(D-1))|[ab]|}}\\
\nonumber
&&\mathrm{exp}\left(i\int-F^{\alpha}\wedge C_{\alpha\beta}\dual F^{\beta}+R_{ab}\wedge\frac{1}{16\pi l_p^2}\varepsilon^{abcd}e_c\wedge e_d-\left(\nabla\phi\right)^A\wedge \left(\dual\nabla\phi\right)_A\right.\\
\nonumber
&&-\frac{i}{6}\varepsilon_{abcd}e^a\wedge e^b\wedge e^c\wedge\left(\left(\bar{\psi}\overleftarrow{\nabla}\right)_A\gamma^d\psi^A-\bar{\psi}_A\gamma^{d}\left(\nabla\psi\right)^A\right)\\
\nonumber
&&-\frac{1}{12}\left.\vphantom{\int}\left(\chi\left(\phi^A\phi_A-v^2\right)^2+Y_{ABC}\bar{\psi}^A\psi^B\phi^C+\frac{\Lambda}{8\pi l_p^2}\right)\varepsilon_{abcd}e^a\wedge e^b\wedge e^c\wedge e^d\right)\\
\nonumber
&&\delta\left(2\pi il_p^2\bar{\psi}_A\gamma_5\gamma^d\psi^A\varepsilon_{abcd}e^b\wedge e^c-\left(\nabla e\right)_a\right)\vphantom{\int} \\
&& F(\phi_k)\,.
\end{eqnarray}
The quantity $D$ (which appears in the exponent of the determinant of the tetrad) is the dimension of the spacetime manifold, $D=4$, so the value of the exponent of the determinant of the tetrad is $N=(|\alpha|+|A|(D-1))|[ab]|=144$, taking into account that $|\alpha| = 12$, $|A| = 4$ and $|[ab]| = 6$. One can recognize that the obtained argument of the exponent is now the action (\ref{eq:ECaction}) of the EC theory, so we can write:
\begin{eqnarray} \label{eq:stepfive}
\nonumber
 Z_{3BF} \langle F \rangle_{3BF} & = & {\cal N}\int De_\partial D\phi_\partial D\psi_\partial D\bar{\psi}_\partial\,\delta(\phi_{\partial})\delta(\psi_\partial)\delta(\bar{\psi}_\partial)\delta(e_\partial)\vphantom{\int} \\
\nonumber
&& \int D\alpha D\omega De D\phi D\psi D\bar{\psi} \; \frac{1}{|e|^N} \; \delta\left(2\pi il_p^2\bar{\psi}_A\gamma_5\gamma^d\psi^A\varepsilon_{abcd}e^b\wedge e^c-\left(\nabla e\right)_a\right)\vphantom{\int}\\
&& F(\phi_k) \; e^{i S_{EC}[\phi_k]}\,. \label{SMsadeltom}
\end{eqnarray}

At this point, one could be tempted to try to establish a correspondence between the quantum $3BF$ theory and the quantum EC theory. Unfortunately, this is not a viable option since the Dirac delta term under the integral enforces an additional strong constraint between torsion and the spin tensor, and must be integrated out before one can attempt to establish the correspondence. This can be done by integrating out the spin connection 1-form $\omega^{[ab]}$, which is present inside the Dirac delta term as part of the covariant derivative $\nabla$ acting on the tetrad 1-form. In order to perform this integration, the expression inside the Dirac delta term must be transformed, since in its original form it depends on the antisymmetric part of the spin connection over the second index and the spacetime index. This dependence can be removed by passing to the locally inertial coordinate system, where this antisymmetric piece can be evaluated, by introducing a change of variables $\omega_{abc}=\omega_{ab\mu}e_{c}{}^\mu$. This change of variables induces the following change of the path integral measure:
\begin{equation}
D\omega_{ab\mu}=D\omega_{abc}\left|\frac{\delta\left(\omega_{abc}e^{c}{}_\mu\right)}{\delta\omega_{efg}}\right|=D\omega_{abc}\left|\delta^{[ef]}_{[ab]}e^{g}{}_\mu\right|=D\omega_{abc}|e|^{|[ab]|}\,.
\end{equation}
Now one can introduce the quantity $A_{abc}$ which is antisymmetric with respect to the second and third indices of the spin connection:
\begin{equation}
A_{abc}=\frac{1}{2}\left(\omega_{abc}-\omega_{acb}\right)\,.
\end{equation}
One can easily demonstrate that these fields contain all components of the spin connection. To see this, it is sufficient to apply the antisymmetry of the spin connection with respect to the first two indices onto the following linear combination:
\begin{equation}
A_{abc}-A_{bac}-A_{cab}=\frac{1}{2}\left(\omega_{abc}-\omega_{acb}-\omega_{bac}+\omega_{bca}-\omega_{cab}+\omega_{cba}\right)=\omega_{abc}\,.
\end{equation}
The Jacobian $\mathcal{J}$ of this change of variables is a constant, and it can be absorbed in the normalization factor ${\cal N}$, so the path integral measure remains essentially the same:
\begin{equation}
\mathcal{N}\int D\omega_{abc}=\mathcal{N}\int |\mathcal{J}| DA_{abc}=\mathcal{N}'\int DA_{abc}\,.
\end{equation}
Substituting this back into (\ref{SMsadeltom}) we obtain the expression which depends on the fields $A_{abc}$:
\begin{eqnarray}
\nonumber
 Z_{3BF} \langle F \rangle_{3BF} & = & {\cal N}\int De_\partial D\phi_\partial D\psi_\partial D\bar{\psi}_\partial\,\delta(\phi_{\partial})\delta(\psi_\partial)\delta(\bar{\psi}_\partial)\delta(e_\partial)\vphantom{\int}\int D\alpha DA De D\phi D\psi D\bar{\psi}\;\frac{|e|^{|[ab]|}}{|e|^N}\\
 &&\delta\left(2\pi il_p^2\bar{\psi}_A\gamma_5\gamma^d\psi^A\varepsilon_{abcd}e^b{}_\mu e^c{}_{\nu}\varepsilon^{\mu\nu\rho\sigma}-\left(\partial_{\mu} e_{a\nu}\right)\varepsilon^{\mu\nu\rho\sigma}+A_{abc}|e|e_d{}^{\rho}e_{e}{}^{\sigma}\varepsilon^{bcde}\right)\vphantom{\int}\\
 \nonumber
 && F(\phi_k) \; e^{i S_{EC}[\phi_k]}\,.
\end{eqnarray}
Applying the functional identities $(\ref{bozonska_delta_int})$ and $(\ref{fermionska_delta_int})$ we then obtain the expression which can be integrated over the fields $A_{abc}$ in a straightforward manner:
\begin{eqnarray}
\nonumber
Z_{3BF} \langle F\rangle_{3BF} & = & {\cal N}\int De_\partial D\phi_\partial D\psi_\partial D\bar{\psi}_\partial\,\delta(\phi_{\partial})\delta(\psi_\partial)\delta(\bar{\psi}_\partial)\delta(e_\partial)\vphantom{\int}\int D\alpha DA De D\phi D\psi D\bar{\psi} \;\frac{|e|^{|[ab]|}}{|e|^N}\\
&&\delta\left(A_{abc}-\left(\frac{1}{2} c_{abc}-2\pi il_p^2\bar{\psi}_A\gamma_5\gamma^d\psi^A\varepsilon_{abcd}\right)\right)\frac{1}{\left|2|e|\varepsilon^{[bc][de]}e_{[d}{}^{[\rho}e_{e]}{}^{\sigma]}\right|^{|a|}} \vphantom{\int}\\
 \nonumber
 && F(\phi_k) \; e^{i S_{EC}[\phi_k]}\,.
\end{eqnarray}
The next step is the evaluation of the determinant of the product of two tetrad fields and the Levi-Civita tensor. One first evaluates the determinants of the Levi-Civita tensor as $\left|\varepsilon^{[ab][cd]}\right|=\left|\varepsilon^{[\mu\nu][\rho\sigma]}\right|=1$, and then one constructs the identity:
\begin{equation}
\frac{1}{|e|^{|[\mu\nu]|}}=\left|\frac{1}{|e|}\varepsilon^{[\mu\nu][\rho\sigma]}\right|=\left|e_a{}^{[\mu}e_b{}^{\nu]}e_c{}^{[\rho}e_d{}^{\sigma]}\varepsilon^{abcd}\right|=\left|2e_{[a}{}^{[\mu}e_{b]}{}^{\nu]}\right|\left|2e_{[c}{}^{[\rho}e_{d]}{}^{\sigma]}\right|\left|\varepsilon^{[ab][cd]}\right|=\left|2e_{[a}{}^{[\mu}e_{b]}{}^{\nu]}\right|^2\,,
\end{equation}
from which it follows that
\begin{equation}\label{detreminanta}
\left|2e_{[a}{}^{[\mu}e_{b]}{}^{\nu]}\right|=\frac{1}{|e|^{\frac{1}{2}|[\mu\nu]|}}=\frac{1}{|e|^{\frac{1}{2}|[ab]|}}\,.
\end{equation}
Using the obtained relation $(\ref{detreminanta})$ the path integral becomes
\begin{eqnarray}
\nonumber
Z_{3BF} \langle F \rangle_{3BF} & = & {\cal N}\int De_\partial D\phi_\partial D\psi_\partial D\bar{\psi}_\partial\,\delta(\phi_{\partial})\delta(\psi_\partial)\delta(\bar{\psi}_\partial)\delta(e_\partial)\vphantom{\int}\int D\alpha DA De D\phi D\psi D\bar{\psi} \label{AnEquationBeforeMdiscussion}\\
&&\delta\left(A_{abc}-\left(\frac{1}{2}c_{abc} -2\pi il_p^2\bar{\psi}_A\gamma_5\gamma^d\psi^A\varepsilon_{abcd}\right)\right) \frac{1}{|e|^{N+|[ab]|\left(\frac{|a|}{2}-1\right)}} \vphantom{\int}\\
 \nonumber
 && F(\phi_k) \; e^{i S_{EC}[\phi_k]}\,.
\end{eqnarray}
In the case of the Standard Model, the exponent of the determinant of the tetrad is given as $M=N+|[ab]|\left(\frac{|a|}{2}-1\right)=N+6=150$, taking into account that $|a|=4$. The integration over the field $A_{abc}$ substitutes the connection $A_{abc}$ in the action with:
\begin{equation}
A_{abc} =\frac{1}{2}c_{abc} -2\pi il_p^2\bar{\psi}_A\gamma_5\gamma^d\psi^A\varepsilon_{abcd}\,,
\end{equation}
which corresponds to the substitution
\begin{equation}
\omega_{ab\mu}=\Delta_{ab\mu}-2\pi il_p^2\bar{\psi}_A\gamma_5\gamma^d\psi^A\varepsilon_{abcd}e^c{}_{\mu}\,,
\end{equation}
which is in turn the previously described procedure of obtaining the action for ECC theory from the action of the EC theory (see equation (\ref{eq:SpinConnectionRicciCoef})). We thus end up with the expression featuring the action of the ECC theory:
\begin{eqnarray}\label{Z3BF}
\nonumber
Z_{3BF} \langle F \rangle_{3BF} &=&{\cal N}\int De_\partial D\phi_\partial D\psi_\partial D\bar{\psi}_\partial\,\delta(\phi_{\partial})\delta(\psi_\partial)\delta(\bar{\psi}_\partial)\delta(e_\partial)\vphantom{\int}\int D\alpha De D\phi D\psi D\bar{\psi}\; \frac{1}{|e|^{M}}\;F(\phi_k) \; e^{iS_{ECC}[\phi_k]}\\
&=&\mathcal{N}' \, Z_{ECC} \left\langle \frac{1}{|e|^M}F\right\rangle_{ECC}\,. \label{eq:AlmostEvaluatedF3bf}
\end{eqnarray}
As a special case, substituting the unit observable, $F(\phi_k) = 1$, one can evaluate the state sum as
\begin{eqnarray}
\nonumber
Z_{3BF}&=&{\cal N}\int De_\partial D\phi_\partial D\psi_\partial D\bar{\psi}_\partial\,\delta(\phi_{\partial})\delta(\psi_\partial)\delta(\bar{\psi}_\partial)\delta(e_\partial)\vphantom{\int}\int D\alpha De D\phi D\psi D\bar{\psi}\;\frac{1}{|e|^{M}}\;e^{iS_{ECC}[\phi_k]}\\
&=&\mathcal{N}'\, Z_{ECC} \left\langle \frac{1}{|e|^M}\right\rangle_{ECC}\,. \label{eq:EvaluationOfZ3bf}
\end{eqnarray}
As another special case, substituting the observable $F(\phi_k) = |e|^M$, one can evaluate the state sum in a different way, as
\begin{eqnarray}
\nonumber
Z_{3BF} \left\langle |e|^M \right\rangle_{3BF} &=&{\cal N}\int De_\partial D\phi_\partial D\psi_\partial D\bar{\psi}_\partial\,\delta(\phi_{\partial})\delta(\psi_\partial)\delta(\bar{\psi}_\partial)\delta(e_\partial)\vphantom{\int}\int D\alpha De D\phi D\psi D\bar{\psi}\;e^{iS_{ECC}[\phi_k]}\\
&=&\mathcal{N}'\, Z_{ECC} \,. \label{eq:EvaluationOfZecc}
\end{eqnarray}
Combining (\ref{eq:EvaluationOfZ3bf}) and (\ref{eq:EvaluationOfZecc}), we obtain the normalization relation
\begin{equation} \label{eq:NormalizationIdentity}
\left\langle|e|^M\right\rangle_{3BF}\left\langle\frac{1}{|e|^M}\right\rangle_{ECC}=1\,.
\end{equation}
As the last step, substituting (\ref{eq:EvaluationOfZ3bf}) and (\ref{eq:EvaluationOfZecc}) back into (\ref{eq:AlmostEvaluatedF3bf}), and remembering the definitions $(\ref{definicijaopservabli})$, we finally obtain the correspondence between the quantum $3BF$ theory and the quantum ECC theory, in the following form:
\begin{equation}\label{veza}
\left\langle F \right\rangle_{3BF}=\frac{\left\langle\frac{1}{|e|^M}F \right\rangle_{ECC}}{\left\langle\frac{1}{|e|^M}\right\rangle_{ECC}}\,,\qquad\left\langle F \right\rangle_{ECC}=\frac{\left\langle|e|^M F\right\rangle_{3BF}}{\left\langle|e|^M\right\rangle_{3BF}}\,.
\end{equation}

The equations (\ref{veza}) are the nonperturbative correspondence between the expectation values of observables in the quantum $3BF$ and quantum ECC theories that we were looking for, and represent the main result of the paper. The existence of this correspondence emphasizes the importance of the $3BF$ theory, since its quantization automatically gives rise to a quantization of the ECC theory, which is a physically relevant model of quantum gravity with matter of the Standard Model. Namely, it can be argued that the explicit construction of path integral for the quantum $3BF$ theory is, all else being equal, easier to perform than the direct construction of the path integral for the quantum ECC theory itself.

\subsection{\label{SubSecBoundary}Boundary conditions}

One technical detail that should be discussed is the presence of the boundary terms in the state sum (\ref{eq:AlmostEvaluatedF3bf}) and eventual dependence of the observable $F$ on values of fields at the boundary of a spacetime manifold (if it features a boundary). As they stand, the boundary terms in (\ref{eq:AlmostEvaluatedF3bf}) feature Dirac delta functions for the boundary values of certain fields, essentially claiming that those fields should vanish at the boundary. While this is not a big issue for fermion fields (since we typically assume them to be localized to some finite region in the manifold bulk anyway), the situation is rather different for the scalar field, and most importantly, the tetrad field. Namely, the scalar field describes the Higgs sector, which is known to have a nonzero vacuum expectation value, which should be constant throughout spacetime. This is in obvious conflict with the statement that it should be zero at the spacetime boundary. Similarly, the tetrad field is typically assumed to be nondegenerate (the tetrad determinant $e$ is assumed to be nonzero everywhere), since otherwise there would be singularities in the metric structure of spacetime. This is again in obvious conflict with the statement that tetrad fields should become zero at the spacetime boundary.

There are two ways one can think of this issue. One way would be to postulate that spacetime does not (or should not) have a boundary to begin with. Then one could simply drop the boundary terms from (\ref{eq:AlmostEvaluatedF3bf}), and the issue of the values of Higgs and tetrad fields at the boundary would become immaterial. Another way would be to modify the $3BF$ action by adding suitable boundary counterterms. These extra terms would not influence the dynamics of the theory in the bulk, but would influence the boundary values of fields. For example, this is a property of the well known Gibbons-Hawking-York (GHY) boundary term \cite{York,GibbonsHawking}. It would thus be interesting to introduce such modifications to the $3BF$ action, and study their influence on the expectation values of observables in (\ref{eq:AlmostEvaluatedF3bf}), with a possibility of fixing the issues related to the Higgs and tetrad fields.

In order to discuss GHY term, we first need some elementary formalism to describe the boundary of the spacetime manifold. Looking at some coordinate patch $x^\mu$ of $\cM$ which intersects the boundary $\del\cM$, one can introduce a new set of coordinates $\xi^i$ ($i = 1,2,3$) on the intersection patch in $\del\cM$. Given this, a point with coordinates $\xi^i$ on $\del\cM$ can be assigned coordinates $x^\mu$ on $\cM$ using the parametric equations
\begin{equation} \label{eq:ParametricBoundaryEq}
x^\mu = z^\mu(\xi^i)\,,
\end{equation}
  where $z^\mu(\xi)$ are some functions encoding the ``position'' of $\del\cM$ in $\cM$. In the tangent space of point $\xi^i$ one can introduce the natural coordinate basis $u_i \equiv \del_i$. These vectors also live in the tangent space of the same point in $\cM$, so one can expand them in the coordinate basis $\del_\mu$ as $u_i = u_i^\mu \del_\mu $, where the components $u_i^\mu $ can be evaluated using (\ref{eq:ParametricBoundaryEq}) as
\begin{equation}
u_i^\mu = \frac{\del z^\mu (\xi)}{\del \xi^i}\,.
\end{equation}
Next, given a metric $g_{\mu\nu}(x)$ on $\cM$, one introduces the induced metric $\gamma_{ij}(\xi)$ on $\del\cM$ as a pullback:
\begin{equation}
\gamma_{ij}(\xi) = g_{\mu\nu}(z(\xi)) \, u_i^\mu u_j^\nu\,.
\end{equation}
This metric (in older literature also called the {\em first fundamental form} on $\del\cM$) is assumed to be nondegenerate, with its inverse denoted as $\gamma^{ij}$, and can be used to raise and lower the boundary indices $i,j,\dots$. One can additionally introduce the induced triads, connection, covariant derivative, curvature, torsion, and various other induced quantities on the boundary, but this is not necessary for the purpose of this work. In addition to all these standard geometric notions associated to $\del\cM$ as a manifold, the boundary has some additional properties that stem from its embedding into $\cM$. Namely, given that $\cM$ is $4$-dimensional and $\del\cM$ is $3$-dimensional, there will be one additional vector in the tangent space of $\cM$ that is linearly independent of the three tangent vectors $u_i$. Calling it the {\em normal vector}, and denoting its components as $n^\mu$, one can choose it to be orthogonal to all three tangent vectors, $n_\mu u^\mu_i =0$, so that the following resolution of the identity holds:
\begin{equation}
\delta^\mu_\nu = \epsilon n^\mu n_\nu + u_i^\mu u^i_\nu\,, \qquad (\epsilon = \pm 1)\,.
\end{equation}
The normal vector is normalized as $n^\mu n_\mu = \epsilon$, and the boundary $\del\cM$ is called spacelike if its normal vector is timelike ($\epsilon = -1$, recall that we work with the $(-,+,+,+)$ signature convention throughout the paper), while it is called timelike if its normal vector is spacelike ($\epsilon = +1$). We will not introduce lightlike boundary since it is not necessary for our purposes.

The normal vector allows us to introduce one more notion specific to the boundary, called {\em extrinsic curvature} (in older literature also called the {\em second fundamental form}), as a projection of the covariant derivative of the normal vector onto the tangent space of $\del\cM$,
\begin{equation}
K_{ij} = u^\mu_i u^\nu_j \nabla_\mu n_\nu\,,
\end{equation}
where $\nabla_\mu$ denotes the standard covariant derivative on $\cM$, compatible with the metric $g_{\mu\nu}$. The extrinsic curvature scalar is defined as $K = \gamma^{ij} K_{ij}$.

Now we are ready to introduce the GHY boundary term. In the context of the traditional Einstein-Hilbert formulation of the action for GR, one can write:
\begin{equation} \label{eq:EHformGHY}
S_{GR} = -\frac{1}{16\pi G} \int_\cM d^4x \; \sqrt{-g} \, R - \frac{1}{8\pi G} \oint_{\del\cM} d^3\xi\; \epsilon \sqrt{|\gamma|} K\,.
\end{equation}
The first term is the standard Einstein-Hilbert action for GR, while the second term is the GHY boundary term. Its purpose is to make sure that the action $S_{GR}$ has well-defined functional derivatives with respect to the metric $g_{\mu\nu}$, since the curvature scalar $R$ contains its second derivatives. Namely, in the variation of the action $\delta S_{GR}$ the second derivatives of the metric in $R$ give rise to the variation of the first derivatives of the metric on the boundary, which are then canceled by the GHY term, rendering $\delta S_{GR}$ ultimately dependent solely on the variation $\delta g_{\mu\nu}$ of the metric itself, rather than its derivatives \cite{York,GibbonsHawking}.

The form of the GHY term in (\ref{eq:EHformGHY}) has been designed precisely to correspond to the Einstein-Hilbert formulation of GR, and does not a priori fulfill its purpose when the gravitational interaction is formulated in terms of the Einstein-Cartan action, or the $3BF$ action, or otherwise. Nevertheless, in \cite{Erdmenger2023} the GHY boundary term has been reformulated to match a completely arbitrary theory that may contain curvature, torsion, and even nonmetricity. In our work, nonmetricity is absent, but curvature and torsion are present, so the results of \cite{Erdmenger2023} lend themselves to be applied in a straightforward manner to the case of the $3BF$ action (\ref{eq:RealisticAction}). Writing (\ref{eq:RealisticAction}) in the form
\begin{equation}
S_{3BF} = \int_\cM \cL_{3BF}\,,
\end{equation}
where $\cL_{3BF}$ is the Lagrangian $4$-form of the $3BF$ action, the GHY term is given as:
\begin{equation}
S_{GHY}^{3BF} = 2\oint_{\del\cM} \epsilon \, K^i \wedge n^\mu u_i^\nu \dual \varphi_{\mu\nu} \Big|_{\del\cM}\,.
\end{equation}
Here $K^i$ is the extrinsic curvature boundary $1$-form
\begin{equation}
  K^i \equiv K^i{}_j \, \rmd \xi^j = u^{i\mu} u^\nu_j \nabla_\mu n_\nu \, \rmd \xi^j \,,
\end{equation}
while $\dual \varphi_{\mu\nu}$ are the $2$-forms obtained from the specific details of the Lagrangian $4$-form $\cL_{3BF}$, see \cite{Erdmenger2023}. For the purpose of our work, we are not interested in the precise form of $\dual \varphi_{\mu\nu}$. We merely need to know that it depends on all fields present in the action (\ref{eq:RealisticAction}), i.e., $\dual \varphi_{\mu\nu}$ is a function over the whole kinematical configuration space (\ref{nedinpolja3bf}) and (\ref{dinpolja3bf}).

At this point we can study the influence of the $GHY$ boundary term on the derivation of our main result (\ref{veza}). We begin by modifying the original $3BF$ action by adding to it the GHY boundary term,
\begin{equation}
S_{3BF} \to S_{3BF} + S_{GHY}^{3BF}\,.
\end{equation}
We then proceed through the calculation described in the previous subsection. Starting from (\ref{eq:initialstep}) with the added GHY boundary term, we proceed to (\ref{eq:stepone}), where the boundary contribution to the path integral (the first row of (\ref{eq:stepone})) now generalizes to:
\begin{equation} \label{eq:NewBoundaryTerm}
\begin{array}{c}
\ds \int D\beta_\partial De_\partial D\tilde{\gamma}_\partial D\gamma_{\partial} D\bar{\gamma}_\partial D\phi_\partial D\psi_\partial D\bar{\psi}_\partial
DB_\partial D\zeta_\partial D\Lambda_\partial
D\alpha_\partial D\omega_\partial  D\tilde{\lambda}_\partial D\lambda_\partial D\bar{\lambda}_\partial DH_\partial DM_\partial \vphantom{\ds\int} \\
\ds e^{iS_{GHY}^{3BF} + i\oint\phi^A_\partial\tilde{\gamma}_{A\partial}+\bar{\psi}_{A\partial}\gamma^A_\partial+\bar{\gamma}_{A\partial}\psi^A_\partial-e_{a\partial}\wedge\beta^a_\partial}\,. \vphantom{\ds\int} \\ 
\end{array}
\end{equation}
The difference from the original boundary contribution in (\ref{eq:stepone}) consists of the fact that the additional $S_{GHY}^{3BF}$ term is present, and since it depends on the full configuration space, we explicitly denote the path integrals over all boundary fields.

The subsequent steps, given by (\ref{eq:steptwo}), (\ref{eq:stepthree}), (\ref{eq:stepfour}), (\ref{eq:stepfive}) all the way to (\ref{Z3BF}), involve integrating out a range of variables living in the bulk, without any contributions to the boundary. This means that the boundary term (\ref{eq:NewBoundaryTerm}) appears in (\ref{Z3BF}) instead of the old one. Note that (\ref{Z3BF}) features also the ECC action (\ref{eq:ECCaction}), which should arguably also be corrected by its own version of the GHY boundary term,
\begin{equation} \label{eq:ECCactionBoundary}
S_{ECC} \to S_{ECC} + S_{GHY}^{ECC}\,,
\end{equation}
where
\begin{equation}
S_{GHY}^{ECC} = 2\oint_{\del\cM} \epsilon \, K^i \wedge n^\mu u_i^\nu \dual \tilde{\varphi}_{\mu\nu} \Big|_{\del\cM}\,.
\end{equation}
Here $\dual \tilde{\varphi}_{\mu\nu}$ are the $2$-forms obtained from the specific details of the ECC Lagrangian $4$-form $\cL_{ECC}$. It is a function over the kinematical configuration space (\ref{dinpoljaecc}) of ECC theory. Subtracting the term $S_{GHY}^{ECC}$ from the boundary term (\ref{eq:NewBoundaryTerm}), and reabsorbing it into a redefinition (\ref{eq:ECCactionBoundary}) of $S_{ECC}$, the final generalized form of equation (\ref{Z3BF}) reads:
\begin{equation}\label{Z3BFboundary}
\begin{array}{lcl}
  Z_{3BF} \langle F \rangle_{3BF} &=&\ds {\cal N} \int D\beta_\partial De_\partial D\tilde{\gamma}_\partial D\gamma_{\partial} D\bar{\gamma}_\partial D\phi_\partial D\psi_\partial D\bar{\psi}_\partial
DB_\partial D\zeta_\partial D\Lambda_\partial
D\alpha_\partial D\omega_\partial  D\tilde{\lambda}_\partial D\lambda_\partial D\bar{\lambda}_\partial DH_\partial DM_\partial \vphantom{\ds\int} \\
 & & \ds \hphantom{mmmmm} e^{iS_{GHY}^{3BF} - iS_{GHY}^{ECC} + i\oint\phi^A_\partial\tilde{\gamma}_{A\partial}+\bar{\psi}_{A\partial}\gamma^A_\partial+\bar{\gamma}_{A\partial}\psi^A_\partial-e_{a\partial}\wedge\beta^a_\partial} \vphantom{\ds\int} \\
& & \ds  \int D\alpha De D\phi D\psi D\bar{\psi}\; \frac{1}{|e|^{M}}\;F(\phi_k) \; e^{iS_{ECC}[\phi_k]} \vphantom{\ds\int} \\
&=& \ds \mathcal{N}' \, Z_{ECC} \left\langle \frac{1}{|e|^M}F\right\rangle_{ECC}\,. \vphantom{\ds\int} \\
\end{array}
\end{equation}
The subsequent analysis leading to (\ref{veza}) remains unchanged --- one evaluates (\ref{Z3BFboundary}) for observables $F(\phi_k) = 1$ and $F(\phi_k) = |e^M|$ and combines them to obtain both (\ref{eq:NormalizationIdentity}) and (\ref{veza}), in an unchanged form, completing the main result.

Looking at the boundary term in (\ref{Z3BFboundary}), we can observe that it is substantially different from the boundary term in (\ref{Z3BF}). Namely, both GHY terms in the exponent depend on all variables in the configuration space, which means that the four remaining terms cannot be integrated in a straightforward manner, and one does not obtain the problematic Dirac delta functions $\delta(\phi_{\partial})\delta(\psi_\partial)\delta(\bar{\psi}_\partial)\delta(e_\partial)$. The evaluation of the explicit form of $S_{GHY}^{3BF}$ and $S_{GHY}^{ECC}$, and the detailed study of its contribution to the boundary integral, are out of the scope of the current work, and we postpone them for future research. Nevertheless, even without explicit calculation, it is reasonably obvious that the GHY terms will give nontrivial contribution and will either completely remove the Dirac delta functions, or substantially modify their arguments, allowing the tetrad, scalar and fermion fields to have potentially nonzero values on the boundary.

\section{\label{secIV}Examples}

In this Section we will compare the predictions of the quantum $3BF$ and $ECC$ theories on the example of the spacetime volume density and discuss the classical limit. Then we will study the example of the gravitational waves.

\subsection{Spacetime 4-volume density}

As a simplest example, let us define the 4-volume density observable defined as
\begin{equation}
F(\phi_k) = \rho \equiv |e|\,,
\end{equation}
and the value of the 4-volume density as the expectation value of this operator, in a given quantum theory of gravity. The name is motivated by the fact that the 4-volume of some 4-dimensional region $\cR$ of spacetime is given as
\begin{equation}
V(\cR) = \int_\cR d^4x\, \sqrt{-g} = \int_\cR d^4x\, |e|\,,
\end{equation}
hence one can loosely call $|e|$ as the ``density'' of spacetime 4-volume of the region $\cR$.

The correspondence relations (\ref{veza}) can be applied to obtain the ratio of the expectation values of 4-volume density in the two quantum theories:
\begin{equation}
\frac{\rho_{3BF}}{\rho_{ECC}}\equiv \frac{\left\langle \rho\right\rangle_{3BF}}{\left\langle \rho\right\rangle_{ECC}}=\frac{\left\langle\frac{1}{|e|^{M-1}}\right\rangle_{ECC}}{\left\langle|e|\right\rangle_{ECC}\left\langle\frac{1}{|e|^{M}}\right\rangle_{ECC}}=\frac{\left\langle|e|\right\rangle_{3BF}\left\langle|e|^{M}\right\rangle_{3BF}}{\left\langle|e|^{M+1}\right\rangle_{3BF}}\,.
\end{equation}
At this point it is useful to remember the definitions of the statistical quantities of covariance and variance, which are useful to separate the quantum corrections from the classical quantities:
\begin{equation}\label{covvar}
\mathrm{Cov}(X,Y)=\langle XY\rangle-\langle X\rangle\langle Y\rangle\,,\qquad\mathrm{Var}(X)=\mathrm{Cov}(X,X)=\left(\Delta X\right)^2\,.
\end{equation}
Here $\Delta X$ represents the standard deviation, i.e., the uncertainty of the observable $X$. Covariance and variance satisfy the Cauchy-Schwarz inequality
\begin{equation} \label{CauchySchwarzInequality}
|\mathrm{Cov}(X,Y)|\leq \Delta X \, \Delta Y\,,
\end{equation}
which can be used to estimate the covariance. Based on the definitions (\ref{covvar}), it is clear that the ratio of the 4-volume densities in $3BF$ and ECC theories is given as:
\begin{equation}
\frac{\rho_{3BF}}{\rho_{ECC}}=1+\frac{\mathrm{Cov}\left(|e|,\frac{1}{|e|^M}\right)_{ECC}}{\left\langle|e|\right\rangle_{ECC}\left\langle\frac{1}{|e|^{M}}\right\rangle_{ECC}}\,,\qquad\frac{\rho_{ECC}}{\rho_{3BF}}=1+\frac{\mathrm{Cov}\left(|e|,|e|^M\right)_{3BF}}{\left\langle|e|\right\rangle_{3BF}\left\langle|e|^{M}\right\rangle_{3BF}}\,.
\end{equation}
Then, using (\ref{CauchySchwarzInequality}), we can write
\begin{equation}
  \frac{\rho_{3BF}}{\rho_{ECC}}\leq 1+
\left(\frac{\Delta |e|}{\left\langle|e|\right\rangle} \;
\frac{\Delta \frac{1}{|e|^M}}{\left\langle\frac{1}{|e|^{M}}\right\rangle}\right)_{ECC}\,,
\qquad
\frac{\rho_{ECC}}{\rho_{3BF}} \leq 1+
\left(\frac{\Delta |e|}{\left\langle|e|\right\rangle} \;
\frac{\Delta |e|^M}{\left\langle|e|^{M}\right\rangle}\right)_{3BF}\,.
\end{equation}
In the classical limit one can assume that the uncertainties tend to zero, and we see that the 4-volume densities have approximately the same value in both theories.

In fact, the above example indicates that the classical limits of the two theories are the same. In order to demonstrate this in full generality, one can repeat the above analysis for the case of an arbitrary observable $F(\phi_k)$. Starting from (\ref{veza}), we have
\begin{equation}
  \frac{\left\langle F\right\rangle_{3BF}}{\left\langle F\right\rangle_{ECC}}=\frac{\left\langle\frac{1}{|e|^M}F\right\rangle_{ECC}}{\left\langle\frac{1}{|e|^M}\right\rangle_{ECC} \left\langle F\right\rangle_{ECC}}\,,
  \qquad
\frac{\left\langle F\right\rangle_{ECC}}{\left\langle F\right\rangle_{3BF}}=\frac{\left\langle|e|^M F\right\rangle_{3BF}}{\left\langle|e|^M\right\rangle_{3BF}\left\langle F\right\rangle_{3BF}}\,.
\end{equation}
The expectation value of the product in the numerator can be removed in favor of covariance using (\ref{covvar}), and the covariance can be estimated using the Cauchy-Schwarz inequality (\ref{CauchySchwarzInequality}), leading us to:
\begin{equation}
\frac{\left\langle F\right\rangle_{3BF}}{\left\langle F\right\rangle_{ECC}} \leq 1+
\left(\frac{\Delta F}{\left\langle F\right\rangle} \;
\frac{\Delta \frac{1}{|e|^M}}{\left\langle\frac{1}{|e|^{M}}\right\rangle}\right)_{ECC}\,,
\qquad
\frac{\left\langle F\right\rangle_{ECC}}{\left\langle F\right\rangle_{3BF}} \leq 1+
\left(\frac{\Delta F }{\left\langle F\right\rangle} \;
\frac{\Delta |e|^M}{\left\langle|e|^{M}\right\rangle}\right)_{3BF}\,.
\end{equation}
In the classical limit we expect the uncertainties of the observables to become negligible, $\Delta F \to 0$, giving us
\begin{equation}
  \frac{\left\langle F\right\rangle_{3BF}}{\left\langle F\right\rangle_{ECC}} \leq 1\,,
\qquad
\frac{\left\langle F\right\rangle_{ECC}}{\left\langle F\right\rangle_{3BF}} \leq 1\,.
\end{equation}
Taken together, these two inequalities enforce the result
\begin{equation}
\left\langle F\right\rangle_{3BF}  = \left\langle F\right\rangle_{ECC} \,,
\end{equation}
claiming that in the classical limit all observables have the same value in both theories, as expected. The two theories differ only at the level of quantum correction terms.

\subsection{\label{subsecGravWaves}Gravitational waves}

In the example of gravitational waves, we study the two (mutually related) fields, namely the tetrad field and the metric field, as well as their excitations over the flat spacetime configuration:
\begin{equation} \label{eq:GWtetradAndMetric}
e^a{}_\mu=\delta^a_{\mu}+\varepsilon^a{}_\mu\,,\qquad g_{\mu\nu}=\eta_{\mu\nu}+h_{\mu\nu}\,.
\end{equation}
We expand the determinant to the second order (which is incidentally the full expansion to all orders):
\begin{equation} \label{eq:GWdet}
\left|e\right|=1+\varepsilon^a{}_a+\frac{1}{2}\left(\varepsilon^a{}_a\varepsilon^b{}_{b}-\varepsilon^a{}_{b}\varepsilon^b{}_a\right)+\frac{1}{6}\left(\varepsilon^a{}_a \varepsilon^b{}_b \varepsilon^c{}_c-3\varepsilon^a{}_a \varepsilon^b{}_c\varepsilon^c{}_b+2\varepsilon^a{}_b\varepsilon^b{}_c\varepsilon^{c}{}_a\right)+|\varepsilon|\,.
\end{equation}
Let us note here that in principle one can discuss gravitational wave perturbations over some more general curved background spacetime, rather than flat spacetime. In other words, given some background tetrad $\hat{e}^a{}_\mu$ and its corresponding background metric $\hat{g}_{\mu\nu}$, instead of (\ref{eq:GWtetradAndMetric}) we could write
\begin{equation}
e^a{}_\mu=\hat{e}^a{}_{\mu}+\varepsilon^a{}_\mu\,,\qquad g_{\mu\nu}=\hat{g}_{\mu\nu}+h_{\mu\nu}\,.
\end{equation}
While the subsequent analysis is in principle similar, it is technically more complicated since the evaluation of the determinant (\ref{eq:GWdet}) would not have $1$ as the leading term, but rather $\det \hat{e}^a{}_\mu$. Because of this, we opt not to work with a generic curved background, but rather with the special case of the flat background. Qualitatively speaking, all results and conclusions of the analysis remain the same.

One can introduce the quantity $E$ that collects all corrections of the tetrad determinant with respect to unity. This is useful for the study of the convergence properties of the power series of $\pm M$-th degree of the tetrad determinant over such a parameter, so the corresponding exponents of the tetrad determinant in second order are given as:
\begin{equation}
\frac{1}{\left|e\right|^{M}}=\frac{1}{(1+E)^M}=\sum_{n=0}^{+\infty}\binom{M+n}{n}\left(-E\right)^n=1-M \varepsilon^a{}_a+\frac{M}{2}\left(M\varepsilon^a{}_a\varepsilon^b{}_{b}+\varepsilon^a{}_{b}\varepsilon^b{}_a\right)+o(\varepsilon^2)\,,
\end{equation}
The requirement that this series converges is that $|E|<1$, while the terms in the series begin to decrease when the following condition is satisfied:
\begin{equation}\label{prvanejednakost}
|E|<\frac{n+1}{M+n+1}\,.
\end{equation}
From here it follows that the contribution of the second order will be greater that the contribution of the third order in case when $|E|<0.0196$ (assuming that $n=2$, $M=150$), or in other words, if one requires that the second order contributes $k$ times more than the third order, then $|E|<0.0196/k$. Also, in the case of the tetrad determinant with a positive exponent, we similarly have:
\begin{equation}
\left|e\right|^M=\left(1+E\right)^M=\sum_{n=0}^{+\infty}\binom{M}{n}E^n=1+M\varepsilon^a{}_a+\frac{M}{2}\left(M\varepsilon^a{}_a\varepsilon^b{}_{b}-\varepsilon^a{}_{b}\varepsilon^b{}_a\right)+o(\varepsilon^2)\,.
\end{equation}
This series is finite, since the binomial coefficient is equal to zero when $n>M$, so there are no convergence issues. In addition, the terms in the series decrease when:
\begin{equation}
|E|<\frac{n+1}{M-n}\,,
\end{equation}
which is a weaker requirement from the previous inequality ($\ref{prvanejednakost}$) and therefore always satisfied.

By applying the correspondence equations ($\ref{veza}$) to the metric perturbation observable, $F(\phi_k) = h_{\mu\nu}$, we obtain the following:
\begin{equation}
\left\langle h_{\mu\nu}\right\rangle_{3BF}=\frac{\left\langle h_{\mu\nu}\right\rangle_{ECC}-M\left\langle \varepsilon^a{}_a h_{\mu\nu}\right\rangle_{ECC}}{1-M\left\langle \varepsilon^a{}_a\right\rangle_{ECC}+\frac{M}{2}\left\langle M\varepsilon^a{}_a\varepsilon^b{}_{b}+\varepsilon^a{}_{b}\varepsilon^b{}_a\right\rangle_{ECC}}\,.
\end{equation}
After the expansion of the denominator into power series one obtains:
\begin{eqnarray}
\nonumber
\left\langle h_{\mu\nu}\right\rangle_{3BF}&=&\left(\left\langle h_{\mu\nu}\right\rangle_{ECC}-M\left\langle \varepsilon^a{}_a h_{\mu\nu}\right\rangle_{ECC}\right)\\
\nonumber
&&\left(1+M\left\langle \varepsilon^a{}_a\right\rangle_{ECC}-\frac{M}{2}\left\langle M\varepsilon^a{}_a\varepsilon^b{}_{b}+\varepsilon^a{}_{b}\varepsilon^b{}_a\right\rangle_{ECC}+\frac{M^2}{2}\left\langle \varepsilon^a{}_a\right\rangle^2_{ECC}\right)\\
&=&\left\langle h_{\mu\nu}\right\rangle_{ECC}\left(1+M\left\langle \varepsilon^a{}_a\right\rangle_{ECC}\right)-M\left\langle \varepsilon^a{}_a h_{\mu\nu}\right\rangle_{ECC}\,.
\end{eqnarray}
Also, in the opposite case we have
\begin{eqnarray}
\nonumber
\left\langle h_{\mu\nu}\right\rangle_{ECC}&=&\left(\left\langle h_{\mu\nu}\right\rangle_{3BF}+M\left\langle \varepsilon^a{}_a h_{\mu\nu}\right\rangle_{3BF}\right)\\
\nonumber
&&\left(1-M\left\langle \varepsilon^a{}_a\right\rangle_{3BF}-\frac{M}{2}\left\langle M\varepsilon^a{}_a\varepsilon^b{}_{b}-\varepsilon^a{}_{b}\varepsilon^b{}_a\right\rangle_{3BF}-\frac{M^2}{2}\left\langle \varepsilon^a{}_a\right\rangle^2_{3BF}\right)\\
&=&\left\langle h_{\mu\nu}\right\rangle_{3BF}\left(1-M\left\langle \varepsilon^a{}_a\right\rangle_{3BF}\right)+M\left\langle \varepsilon^a{}_a h_{\mu\nu}\right\rangle_{3BF}\,.
\end{eqnarray}
These expressions simplify to the following final form:
\begin{eqnarray}
\left\langle h_{\mu\nu}\right\rangle_{3BF}=\left\langle h_{\mu\nu}\right\rangle_{ECC}-M\mathrm{Cov}\left(\varepsilon^a{}_a\,, h_{\mu\nu}\right)_{ECC}\,, \label{eq:vezaEcciBFjedna}\\
\left\langle h_{\mu\nu}\right\rangle_{ECC}=\left\langle h_{\mu\nu}\right\rangle_{3BF}+M\mathrm{Cov}\left(\varepsilon^a{}_a\,, h_{\mu\nu}\right)_{3BF}\,. \label{eq:vezaEcciBFdruga}
\end{eqnarray}

Using the Cauchy-Schwarz inequality (\ref{CauchySchwarzInequality}), we can estimate the order of magnitude of the perturbation necessary for the experimental comparison between quantum $3BF$ and quantum ECC theories. Namely, up to second order, the deviation of the predictions between two theories is given as
\begin{equation}\label{cov}
\mathrm{Cov}\left(\varepsilon^a{}_a\,, h_{\mu\nu}\right)_{ECC}=\mathrm{Cov}\left(\varepsilon^a{}_a\,, h_{\mu\nu}\right)_{3BF}=\mathrm{Cov}\left(\varepsilon^a{}_a\,, h_{\mu\nu}\right)\,,
\end{equation}
i.e., the correction term is the same in both theories up to second order, which can be seen from (\ref{eq:vezaEcciBFjedna}) and (\ref{eq:vezaEcciBFdruga}). Besides, based on the relations $h_{\mu\nu}=\eta_{\mu a}\varepsilon^a{}_{\nu}+\eta_{a\nu}\varepsilon^a{}_{\mu}$, $\varepsilon_{\mu\nu}=\eta_{\mu a}\varepsilon^a{}_{\nu}$ and the Cauchy-Schwarz inequality, assuming also that the uncertainty of each component of the tetrad is approximately the same, one obtains that:
\begin{equation} \label{GWinequality}
\left\langle h_{\mu\nu}\right\rangle_{ECC}-\left\langle h_{\mu\nu}\right\rangle_{3BF}=M\mathrm{Cov}\left(\varepsilon^a{}_a\,, h_{\mu\nu}\right)\leq 2M \Delta\varepsilon^a{}_a \, \Delta \varepsilon_{\mu\nu} \approx 8M \left( \Delta \varepsilon_{\mu\nu}\right)^2 \approx 2M\left(\Delta h_{\mu\nu}\right)^2\,.
\end{equation}

The inequality (\ref{GWinequality}) can in principle be used to experimentally distinguish between the quantum $3BF$ theory and quantum ECC theory, by measuring the gravitational waves and comparing the outcome to theoretical predictions. In order to obtain some intuition of the orders of magnitude involved, let us start from some ballpark orders of magnitude for current technological state of the art measurements of gravitational waves, taking for example LIGO/Virgo detectors as reference. According to \cite{LIGOpaper}, a typical precision of the strain measurement can be estimated to be $10^{-21}$, which means that the right-hand side of (\ref{GWinequality}) should be
\begin{equation}
2M\left(\Delta h_{\mu\nu}\right)^2 \geq 10^{-21}\,.
\end{equation}
Remembering that for the Standard Model we have $M=150$ (see discussion below equation (\ref{AnEquationBeforeMdiscussion})), this gives us an estimate for the minimal quantum correction that can be detectable:
\begin{equation}
\Delta h_{\mu\nu} \geq \sqrt{\frac{10^{-21}}{2\cdot 150}} \approx 10^{-12}\,.
\end{equation}
This is a huge value, as can be seen from the fact that the strain amplitude of the black hole merger signal in \cite{LIGOpaper} is of order $10^{-18}$. While the distance of GW150914 source was estimated to $r_{GW} \approx 410\, {\rm Mpc}$, which is far outside of our galaxy, one can infer the strain amplitude of a hypothetical similar event happening within the Milky Way galaxy, i.e., at a distance of $r_{MW} \approx 34\, {\rm Kpc}$. Since the amplitude of the strain of a spherical wave falls off as $1/r$ from the source, one could simply estimate that a similar black hole merger within our galaxy would give rise to the signal with strain amplitude of the order
\begin{equation}
h_{MW} \approx h_{GW} \frac{r_{GW}}{r_{MW}} = 10^{-18} \times \frac{ 4.1 \cdot 10^{5} \, {\rm Kpc} }{3.4 \cdot 10^{1} \, {\rm Kpc}} \approx 10^{-14}\,.
\end{equation}
This is still two orders of magnitude smaller than the needed scale of the quantum correction $\Delta h_{\mu\nu}$. Moreover, there is nothing in the theory to suggest why a system of two merging black holes would even have a quantum uncertainty that big, to begin with.

In other words, using current technology, one would need a gravitational wave source that
\begin{itemize}
\item[(a)] generates strain $\left\langle h_{\mu\nu}\right\rangle $ of the order of at least $10^{-11}$, and
\item[(b)] gives rise to quantum uncertainty of the strain, $\Delta h_{\mu\nu} $, of the order of at least $10^{-12}$.
\end{itemize}
Obviously, there are no known candidates for such a source of gravitational waves in nature. Nevertheless, at least in principle, if one were to have such a source, it would be possible to apply (\ref{GWinequality}) to experimentally distinguish between the quantum $3BF$ theory and the quantum ECC theory.

\section{\label{secV}Conclusions}

\subsection{\label{secVa}Summary of the results}

Let us summarize the results of the paper. After the Introduction, in Section \ref{secII} we gave a short review of four classical actions --- the topological $3BF$ action, the physically relevant constrained $3BF$ action, the Einstein-Cartan action featuring the Standard Model matter sector, and the Einstein-Cartan contact action, featuring the four-fermion contact interaction. We have demonstrated that the constrained $3BF$ theory and the ECC theory give rise to equivalent sets of classical equations of motion. In Section \ref{secIII}, we have turned to the main analysis of the expectation values of an arbitrary quantum observable $F(\phi_k)$ that can be defined in both theories. After introducing some mathematical identities needed for the analysis, we have established a correspondence between the expectation value of the observable $F$ in one theory, and the expectation value of a corresponding observable $|e|^{\pm M} F$ in the other theory, where $e$ is the determinant of the tetrad fields, while the coefficient $M$ has been determined to be $M=150$. This correspondence has been established in a fully nonperturbative way, and represents the main result of the paper. Section \ref{secIII} closes with an analysis of the boundary terms present in the theory. In Section \ref{secIV} we discussed some illustrative example observables, in order to compare the two quantum theories. First, we have discussed the spacetime 4-volume density as a simple example, and also the classical limit of the two theories. Then, we turned our attention to the example of gravitational waves, and we gave an estimate of how large their quantum uncertainties must be in order to be able to experimentally distinguish between the two quantum theories. 

\subsection{\label{secVb}Discussion}

The main relevance of our results is reflected in the following. On one hand, the classical ECC theory is arguably the phenomenologically very relevant model for the construction of a realistic full theory of quantum gravity with matter. Needless to say, the quantization of this action is rather hard, and so far remains an open problem in modern theoretical physics. On the other hand, the classical constrained $3BF$ theory represents a model that is slightly more tangible for efficient and rigorous quantization, at least within the path integral formalism. This is partly because it is based on a somewhat novel algebraic structure, a $3$-group, which represents the generalization of the notion of a Lie group within the framework of higher gauge theory. Initial steps towards the path integral quantization of the $3BF$ theory for the case of the Standard Model $3$-group have already been taken \cite{Radenkovic2019,Radenkovic2020,Radenkovic2022a,Radenkovic2020proc,Radenkovic2022b,Djordjevic2023,Stipsic2025,Radenkovic2025}, with promising results. Given this context, establishing a fully nonperturbative correspondence between the quantum $3BF$ and quantum ECC theory represents a quite useful result, since it allows us to sidestep the hard question of quantization of ECC theory itself, and instead define it in terms of the quantization of the $3BF$ theory.

The second important consequence of the correspondence is that there exists a regime where the $3BF$ and ECC theories could be experimentally distinguished from each other, at least in principle. As we have seen in Subsection \ref{subsecGravWaves}, given a source of gravitational waves that is both strong in magnitude and has large quantum uncertainty, the difference in the quantum corrections for the strain amplitude can in principle be large enough to be detected using current technology. This would allow us to compare $3BF$ and ECC theories against experimental data. In this sense, the correspondence predicts observable signatures that distinguish the two theories. Obviously, we do not have actual access to a gravitational wave source with the required properties, so any such experimental proposal is outside of the realm of practical feasibility. But as a matter of principle,  the fact that such scenarios can be studied at least theoretically illustrates the phenomenological significance of the obtained correspondence.

Going beyond the $3BF$ and ECC theories, one can also ask is it possible to establish a similar correspondence in the context of other approaches to quantum gravity, such as the canonical loop quantum gravity (LQG), or causal set theory (CST)? This is an interesting question, with no obvious answer. Namely, the correspondence between $3BF$ and ECC theories has been established using the language of path integrals. In the canonical LQG approach one starts essentially from a variant of EC theory, but the path integal language is not used. Instead, the quantization is being performed by foliating the spacetime into space and time, and by imposing canonical commutation relations on appropriate variables on each spatial hypersurface \cite{Rovelli2004,Thiemann}. This canonical quantization programme has not been developed for the $3BF$ theory so far. Nevertheless, assuming it could be developed, one could in principle study the same correspondence within the canonical language --- given an observable in the canonical quantization of $3BF$ theory, can one find a corresponding observable in canonical LQG, such that their expectation values are equal? This may be an interesting topic for future research.

Regarding the possible correspondence between $3BF$ theory and causal set theory, the situation is more complicated. While CST can be formulated using the (discrete) path integral language, there are two main issues that arise when trying to establish the correspondence between observables in $3BF$ and CST. First, the CST formalism has so far not been developed enough to describe non-scalar matter fields \cite{Surya2019}. This is problematic, since the $3BF$ theory features gravity coupled to the full Standard Model, including fermions and gauge bosons. Therefore, a correspondence between most of the observables in $3BF$ and CST theories is impossible to establish. Second, in the classical limit, the $3BF$ theory gives rise to standard Einstein field equations of general relativity. On the other hand, the classical limit of CST cannot reproduce full Einstein field equations, since the fundamental CST assumption of a causal order relation excludes some of their solutions. For example, closed timelike curves are well known solutions of Einstein equations, but cannot be present within the CST formalism because they describe spacetime geometry with causal structure containing cycles, which is incompatible with a poset-type order relation required by CST. In this sense, $3BF$ and CST theories have different classical limits (more precisely, their sets of classical solutions are inequivalent), suggesting that the general correspondence between observables of the two theories is unlikely to exist.

When looking at the procedure of deriving the correspondence relations (\ref{veza}) in Section \ref{secIII}, there are some technical details that merit further attention. Namely, the relation (\ref{SMsadeltom}) comes tantalizingly close to establishing a correspondence between the $3BF$ theory and the standard Einstein-Cartan theory, as opposed to the ECC theory. As was discussed below (\ref{SMsadeltom}), such a prospect is thwarted due to the presence of the Dirac delta function encoding the algebraic relationship between the spin connection and fermion spin current, which is a consequence of the typical coupling between fermions and torsion in EC theory. The constrained $3BF$ action (\ref{eq:RealisticAction}) has been constructed to encode precisely that same coupling, so the presence of the corresponding Dirac delta function should come as no surprise. Nevertheless, it would be an interesting avenue of research to extend both the $3BF$ theory and the EC theory to include some different type of coupling between fermions and torsion, which could potentially lead to a proper dynamical equation of motion for the spin connection. This is a valid possibility, since phenomenologically nobody in fact knows precisely how fermions couple to torsion --- this interaction (if it exists to begin with) is too small and has so far eluded any experimental detection. In this context, one could formulate some alternative $3BF$ and EC models, which feature a different type of fermion-torsion coupling. Then one could attempt to repeat the analysis presented in this paper for those models, possibly establishing the correspondence between the quantum versions of the modified $3BF$ theory and the modified EC theory, rather than the ECC theory. This seems to be an interesting topic for future investigation.

\acknowledgments

The authors would like to thank Tijana Radenkovi\'c and Mihailo \Dj or\dj evi\'c for useful discussions and suggestions.

\medskip

This research was supported by the Ministry of Education, Science and Technological Development of the Republic of Serbia (MPNTR). MV was additionally supported by the Science Fund of the Republic of Serbia, grant number 7745968, project ``Quantum Gravity from Higher Gauge Theory 2021'' -- QGHG-2021. The contents of this publication are the sole responsibility of the authors and can in no way be taken to reflect the views of the Science Fund of the Republic of Serbia.

\appendix

\section{\label{AppA}Proofs of the identities featuring Dirac delta functions}

The proofs of the identities featuring Dirac delta functions for real and Grassmann fields can be established by the following straightforward computations:
\begin{itemize}
\item the identity $(\ref{bozonska_delta_int})$:
\begin{eqnarray}
\nonumber
&&\int D\varphi D\phi_k\,\delta\left(\varphi^{aB}F_{B}{}^{A}(\phi_k)-G^{aA}(\phi_k)\right)H(\varphi,\phi_k)\\
\nonumber
&&=\int D\left(\varphi^{aB}F_{B}{}^{A}(\phi_k)\right)D\phi_k\begin{vmatrix}
\frac{\delta\left(\varphi F(\phi_k)F^{-1}(\phi_k)\right)^{aA}}{\delta\phi_n} & \frac{\delta\phi_m}{\delta\phi_n}\\
\frac{\delta\left(\varphi F(\phi_k)F^{-1}(\phi_k)\right)^{aA}}{\delta\left(\varphi F(\phi_k)\right)^{bB}} & 0
\end{vmatrix}\delta\left(\varphi^{aB}F_B{}^{A}(\phi_k)-G^{aA}(\phi_k)\right)H(\varphi,\phi_k)\\
\nonumber
&&=\int D\hat{\varphi} D\phi_k\left|\delta_{b}^a F^{-1}{}_{B}{}^A(\phi_k)\right|\delta\left(\hat{\varphi}^{aA}-G^{aA}(\phi_k)\right)H\left(\hat{\varphi}^{aB}F^{-1}{}_{B}{}^A(\phi_k),\phi_k\right)\\
&&=\int D\phi_k \frac{1}{\left|F(\phi_k)\right|^{|a|}}H\left(G^{aB}(\phi_k)F^{-1}{}_B{}^A(\phi_k),\phi_k\right)\,,
\end{eqnarray}
\item the identity $(\ref{fermionska_delta_def})$:
\begin{eqnarray}
\nonumber
&&\int_{\grasmanovi^{n}} d\theta_1 d\theta_2 d\theta_3...d\theta_n e^{i\theta_1\left(\theta_2-\theta_3-...-\theta_k\right)}F(\theta_2,\theta_3,...\,,\theta_n)\\
\nonumber
&&=\int_{\grasmanovi^{n}} d\theta_1 d\theta_2 d\theta_3...d\theta_n\left(1+i\theta_1\left(\theta_2-\theta_3-...-\theta_k\right)\right)\\
\nonumber
&&\qquad\times\left(\theta_3...\theta_kf_{01...1}(\theta_{k+1},...\,,\theta_n)+...+\theta_2...\theta_{k-1}f_{1...10}(\theta_{k+1},...\,,\theta_n)\right)\\
\nonumber
&&=i\int_{\grasmanovi^{n}} d\theta_1 d\theta_2 d\theta_3...d\theta_n\,\theta_1\theta_2\theta_3...\theta_k\left(f_{01...1}(\theta_{k+1},...\,,\theta_n)+...+(-1)^{l}f_{11...10_l1...1}(\theta_{k+1},...\,,\theta_n)\right)\\
\nonumber
&&=-i\int_{\grasmanovi^{n}}  d\theta_3...d\theta_nd\theta_2 d\theta_1\,\theta_1\theta_2\theta_3...\theta_k\left(f_{01...1}(\theta_{k+1},...\,,\theta_n)+...+(-1)^{l}f_{11...10_l1...1}(\theta_{k+1},...\,,\theta_n)\right)\\
\nonumber
&&=-i\int_{\grasmanovi^{n-2}}  d\theta_3...d\theta_n\,\theta_3...\theta_k\left(f_{01...1}(\theta_{k+1},...\,,\theta_n)+...+(-1)^{l}f_{11...10_l1...1}(\theta_{k+1},...\,,\theta_n)\right)\\
\nonumber
&&=-i\int_{\grasmanovi^{n-2}}  d\theta_3...d\theta_n\,F(\theta_3+...+\theta_k,\theta_3,...\,,\theta_n) \\
\nonumber
&&=-i\int_{\grasmanovi^{n-1}}  d\theta_3...d\theta_n d\theta_2\,\delta(\theta_2-\theta_3-...-\theta_k)F(\theta_2,\theta_3,...\,,\theta_n)\\
&&=(-1)^{n-1} i\int_{\grasmanovi^{n-1}} d\theta_2...d\theta_n\,\delta(\theta_2-\theta_3-...-\theta_k)F(\theta_2,\theta_3,...\,,\theta_n)\,,
\end{eqnarray}
\item the identity $(\ref{fermionska_delta_int})$:
\begin{eqnarray}
\nonumber
&&\int_{\realni^m}d^m y \int_{\realni^m}d^m x\int_{\grasmanovi^k} d\theta_1 d\theta_2...d\theta_k  \int_{\grasmanovi^{n-k}} d\theta_{k+1}...d\theta_n\,e^{iy_a\left(x^a-M^{aij}\theta_{i}\theta_j\right)}F(x,\theta_1,...\,,\theta_n)\\
\nonumber
&&=\int_{\realni^m}d^m y \int_{\realni^m}d^m x \int_{\grasmanovi^k} d\theta_1 d\theta_2...d\theta_k \int_{\grasmanovi^{n-k}} d\theta_{k+1}...d\theta_n\,e^{iy_a x^a}\sum_{b=0}^{+\infty}\frac{1}{b!}\left(-iy_a M^{aij}\theta_{i}\theta_{j}\right)^bF(x,\theta_1,...\,,\theta_n)\\
\nonumber
&&=\int_{\realni^m}d^m y \int_{\realni^m}d^m x \int_{\grasmanovi^{k}} d\theta_1 d\theta_2...d\theta_k \int_{\grasmanovi^{n-k}} d\theta_{k+1}...d\theta_n\,e^{iy_a x^a}\sum_{b=0}^{+\infty}\frac{1}{b!}\left(M^{aij}\theta_{i}\theta_{j}\frac{\partial}{\partial x^a}\right)^b F(x,\theta_1,...\,,\theta_n)\\
\nonumber
&&=(2\pi)^m\int_{\realni^m}d^m x\int_{\grasmanovi^{k}} d\theta_1 d\theta_2...d\theta_k  \int_{\grasmanovi^{n-k}} d\theta_{k+1}...d\theta_n \prod_{a=1}^m\delta(x^a)\sum_{b=0}^{+\infty}\frac{1}{b!}\left(M^{aij}\theta_{i}\theta_{j}\frac{\partial}{\partial x^a}\right)^b F(x,\theta_1,...\,,\theta_n)\\
\nonumber
&&=(2\pi)^m\int_{\grasmanovi^k} d\theta_1 d\theta_2...d\theta_k\int_{\grasmanovi^{n-k}} d\theta_{k+1}...d\theta_n\,\sum_{b=0}^{+\infty}\frac{1}{b!}\left(M^{aij}\theta_{i}\theta_{j}\frac{\partial}{\partial x^a}\right)^b F(x,\theta_1,...\,,\theta_n)\Big|_{x=0}\\
\nonumber
&&=(2\pi)^m\int_{\grasmanovi^k} d\theta_1 d\theta_2...d\theta_k\int_{\grasmanovi^{n-k}} d\theta_{k+1}...d\theta_n\, F\left(M^{aij}\theta_{i}\theta_{j},\theta_1,...\,,\theta_n\right)\\
&&=(2\pi)^m\int_{\realni^m}d^m x\int_{\grasmanovi^{k}} d\theta_1 d\theta_2...d\theta_k\int_{\grasmanovi^{n-k}} d\theta_{k+1}...d\theta_n\prod_{a=1}^m \delta\left(x^a-M^{aij}\theta_{i}\theta_{j}\right) F(x,\theta_1,...\,,\theta_n)\,.
\end{eqnarray}
\end{itemize}

\section{\label{AppB}Component form of the notation}

Let us illustrate some details regarding the notation introduced in Subsection \ref{subsecIIa}. Given a Lie 2-crossed module (\ref{eq:jna5}),
\begin{equation}
(\; L\stackrel{\delta}{\to} H \stackrel{\partial}{\to}G \; , \;\triangleright \;, \; \{\_\,,\_\,\}_\mathrm{pf} \;)\,,
\end{equation}
the Lie groups $G$, $H$ and $L$ have their corresponding Lie algebras $\mathfrak{g}$, $\mathfrak{h}$ and $\mathfrak{l}$, which allow us to introduce a corresponding linear structure, called a differential Lie 2-crossed module, as
\begin{equation}
(\; \mathfrak{l}\stackrel{\delta}{\to} \mathfrak{h} \stackrel{\partial}{\to}\mathfrak{g} \; , \;\triangleright \;, \; \{\_\,,\_\,\}_\mathrm{pf} \;)\,,
\end{equation}
with the maps
\begin{equation}
\partial : \mathfrak{h} \to \mathfrak{g}\,, \qquad \delta : \mathfrak{l} \to \mathfrak{h}\,,
\end{equation}
\begin{equation}
\triangleright : \mathfrak{g} \times \mathfrak{a} \to \mathfrak{a}\,, \qquad \mathfrak{a} = \mathfrak{g}, \mathfrak{h}, \mathfrak{l}\,,
\end{equation}
\begin{equation}
\{ \_ \, , \_\, \}_\mathrm{pf} : \mathfrak{h} \times \mathfrak{h} \to \mathfrak{l}\,.
\end{equation}
These maps are linearized versions of (\ref{eq:jna2}), (\ref{eq:jna3}) and (\ref{eq:jna4}), and are subject to axioms which are naturally induced by the axioms of a Lie 2-crossed module. 

One can introduce sets of basis vectors for the three algebras $\mathfrak{g}$, $\mathfrak{h}$ and $\mathfrak{l}$, denoted respectively as
\begin{equation}
  \{\,\tau_\alpha \in \mathfrak{g}\, | \, \alpha=1,\dots,\dim\mathfrak{g}\,\}\,, \qquad
  \{\,t_a \in \mathfrak{h}\, | \, a=1,\dots,\dim\mathfrak{h}\,\}\,, \qquad
  \{\,T_A \in \mathfrak{l}\, | \, A=1,\dots,\dim\mathfrak{l}\,\}\,.
\end{equation}
This allows us to introduce the components of the above maps as:
\begin{equation}
\partial t_a = \partial_a{}^\alpha\tau_\alpha\,, \qquad
\delta T_A = \delta_A{}^a t_a\,, \qquad
\{ t_a \, , t_b\, \}_\mathrm{pf} = X_{ab}{}^A T_A\,,
\end{equation}
\begin{equation}
\tau_\alpha \triangleright \tau_\beta  = \triangleright_{\alpha\beta}{}^\gamma \tau_\gamma\,, \qquad
\tau_\alpha \triangleright t_a  = \triangleright_{\alpha a}{}^b t_b\,, \qquad
\tau_\alpha \triangleright T_A  = \triangleright_{\alpha A}{}^B T_B\,.
\end{equation}
Next, given a $4$-dimensional manifold $\cM$, one can denote the space of differential $p$-forms over $\cM$ as $\Lambda_p(\cM)$, and its natural basis as $\rmd x^{\mu_1} \wedge \dots \wedge \rmd x^{\mu_p}$. One can then introduce a principal $3$-group bundle over a $4$-dimensional manifold $\cM$. As a section of this bundle one can introduce the 3-connection $(\alpha, \beta, \gamma)$, which consists of a $\mathfrak{g}$-valued 1-form $\alpha$, an $\mathfrak{h}$-valued 2-form $\beta$ and an $\mathfrak{l}$-valued 3-form $\gamma$:
\begin{equation}
\alpha \in \Lambda_1(\cM)\otimes \mathfrak{g}\,, \qquad
\beta \in \Lambda_2(\cM)\otimes \mathfrak{h}\,, \qquad
\gamma \in \Lambda_3(\cM)\otimes \mathfrak{l}\,.
\end{equation}
These can be expanded into components using appropriate basis vectors as:
\begin{equation}
\alpha = \alpha^\alpha{}_\mu(x) \; \rmd x^\mu \otimes \tau_\alpha\,, \qquad
\beta = \frac{1}{2} \beta^a{}_{\mu\nu}(x) \; \rmd x^\mu \wedge \rmd x^\nu \otimes t_a\,, \qquad
\gamma =\frac{1}{3!} \gamma^A{}_{\mu\nu\lambda}(x) \; \rmd x^\mu \wedge \rmd x^\nu \wedge \rmd x^\lambda \otimes T_A\,.
\end{equation}
The field $\alpha^\alpha{}_\mu(x)$ is precisely a traditional connection of a principal $G$-bundle, while $\beta^a{}_{\mu\nu}(x)$ and $\gamma^A{}_{\mu\nu\lambda}(x)$ are additional fields, native to the framework based on a principal $3$-group bundle.

Given a $3$-connection $(\alpha,\beta,\gamma)$, one can introduce a so-called fake $3$-curvature $(\cF, \cG, \cH)$, consisting of a $\mathfrak{g}$-valued 2-form $\cF$, an $\mathfrak{h}$-valued 3-form $\cG$ and an $\mathfrak{l}$-valued 4-form $\cH$:
\begin{equation}
\cF \in \Lambda_2(\cM)\otimes \mathfrak{g}\,, \qquad
\cG \in \Lambda_3(\cM)\otimes \mathfrak{h}\,, \qquad
\cH \in \Lambda_4(\cM)\otimes \mathfrak{l}\,.
\end{equation}
These are defined as (see equation (\ref{eq:ThreeCurvatureDef}) in the main text):
\begin{equation} \label{eq:ThreeCurvatureDefOpet}
{\cal F}={\rm d}\alpha+\alpha\wedge \alpha-\partial\beta\,,\qquad
{\cal G}={\rm d}\beta+\alpha\wedge^{\triangleright}\beta-\delta\gamma\,,\qquad
{\cal H}={\rm d}\gamma+\alpha\wedge^{\triangleright}\gamma+\{\beta\wedge\beta\}_{\rm pf}\,.
\end{equation}
The notation $\wedge^\triangleright $ means that one should apply the wedge-product $\wedge $ in the subspace of differential forms, while simultaneously apply the action $\triangleright $ in the algebra subspace. One can use the above equations to work out the explicit components of the fake $3$-curvature. For example, the components of the 3-form $\cG$ are obtained as follows. The $\rmd \beta$ term is:
\begin{equation} \label{eq:dbetaterm}
\begin{array}{lcl}
{\rm d}\beta & = & \ds {\rm d}\left( \frac{1}{2} \beta^b{}_{\mu\nu} \; \rmd x^\mu \wedge \rmd x^\nu \otimes t_b \right) \vphantom{\ds\int} \\
 & = & \ds  \frac{1}{2} \del_\lambda \beta^b{}_{\mu\nu} \; \rmd x^\lambda \wedge \rmd x^\mu \wedge \rmd x^\nu \otimes t_b \vphantom{\ds\int} \\
 & = & \ds  \frac{1}{6} \Big[ \del_\lambda \beta^b{}_{\mu\nu} + \del_\mu \beta^b{}_{\nu\lambda} + \del_\nu \beta^b{}_{\lambda\mu} \Big] \; \rmd x^\lambda \wedge \rmd x^\mu \wedge \rmd x^\nu \otimes t_b \,. \vphantom{\ds\int} \\
\end{array}
\end{equation}
The $\alpha \wedge^\triangleright \beta$ term is:
\begin{equation} \label{eq:alphabetaterm}
\begin{array}{lcl}
\alpha\wedge^{\triangleright}\beta & = & \ds \Big( \alpha^\alpha{}_\lambda \; \rmd x^\lambda \otimes \tau_\alpha \Big)\wedge^{\triangleright}\left(\frac{1}{2} \beta^a{}_{\mu\nu} \; \rmd x^\mu \wedge \rmd x^\nu \otimes t_a\right) \vphantom{\ds\int} \\
 & = & \ds \frac{1}{2} \alpha^\alpha{}_\lambda \beta^a{}_{\mu\nu}  \; \left(  \rmd x^\lambda \wedge \rmd x^\mu \wedge \rmd x^\nu \right)  \otimes \left(  \tau_\alpha \triangleright t_a\right) \vphantom{\ds\int} \\
 & = & \ds \frac{1}{2} \triangleright_{\alpha a}{}^b \alpha^\alpha{}_\lambda \beta^a{}_{\mu\nu}  \; \rmd x^\lambda \wedge \rmd x^\mu \wedge \rmd x^\nu \otimes t_b \vphantom{\ds\int} \\
& = & \ds \frac{1}{6} \Big[ \triangleright_{\alpha a}{}^b \alpha^\alpha{}_\lambda \beta^a{}_{\mu\nu} + \triangleright_{\alpha a}{}^b \alpha^\alpha{}_\mu \beta^a{}_{\nu\lambda} + \triangleright_{\alpha a}{}^b \alpha^\alpha{}_\nu \beta^a{}_{\lambda\mu} \Big] \; \rmd x^\lambda \wedge \rmd x^\mu \wedge \rmd x^\nu \otimes t_b \vphantom{\ds\int} \\
\end{array}
\end{equation}
The $\delta \gamma$ term is:
\begin{equation}
\begin{array}{lcl}
\delta\gamma & = & \ds \delta\left(\frac{1}{3!} \gamma^A{}_{\mu\nu\lambda} \; \rmd x^\mu \wedge \rmd x^\nu \wedge \rmd x^\lambda \otimes T_A\right) \vphantom{\ds\int} \\
 & = & \ds \frac{1}{6} \gamma^A{}_{\mu\nu\lambda} \; \rmd x^\mu \wedge \rmd x^\nu \wedge \rmd x^\lambda \otimes \delta T_A \vphantom{\ds\int} \\
 & = & \ds \frac{1}{6} \Big[ \delta_A{}^b \gamma^A{}_{\mu\nu\lambda} \Big] \; \rmd x^\mu \wedge \rmd x^\nu \wedge \rmd x^\lambda \otimes t_b \vphantom{\ds\int} \\
\end{array}
\end{equation}
Putting all three terms together, and comparing to the expansion of $\cG$ into components,
\begin{equation}
\cG =\frac{1}{3!} \cG^b{}_{\lambda\mu\nu}(x) \; \rmd x^\lambda \wedge \rmd x^\mu \wedge \rmd x^\nu \otimes t_b\,,
\end{equation}
one obtains:
\begin{equation}
\cG^b{}_{\lambda\mu\nu} = \del_\lambda \beta^b{}_{\mu\nu} + \del_\mu \beta^b{}_{\nu\lambda} + \del_\nu \beta^b{}_{\lambda\mu} + \triangleright_{\alpha a}{}^b \alpha^\alpha{}_\lambda \beta^a{}_{\mu\nu} + \triangleright_{\alpha a}{}^b \alpha^\alpha{}_\mu \beta^a{}_{\nu\lambda} + \triangleright_{\alpha a}{}^b \alpha^\alpha{}_\nu \beta^a{}_{\lambda\mu} - \delta_A{}^b \gamma^A{}_{\mu\nu\lambda}\,.
\end{equation}
In a similar fashion, one can derive the components for $\cF$ and $\cH$ as well. The result is
\begin{equation}
\cF^\alpha{}_{\mu\nu} = \del_\mu \alpha^\alpha{}_\nu - \del_\mu \alpha^\alpha{}_\nu + \triangleright_{\beta\gamma}{}^\alpha \alpha^\beta{}_\mu \alpha^\gamma{}_\nu - \partial_a{}^\alpha \beta^a{}_{\mu\nu} \,,
\end{equation}
and
\begin{equation}
\begin{array}{lcl}
  \cH^A{}_{\mu\nu\rho\sigma} & = & \ds \del_\mu \gamma^A{}_{\nu\rho\sigma} - \del_\nu \gamma^A{}_{\rho\sigma\mu} + \del_\rho \gamma^A{}_{\sigma\mu\nu} - \del_\sigma \gamma^A{}_{\mu\nu\rho} \vphantom{\ds\int} \\
  & & + \triangleright_{\alpha B}{}^A  \alpha^\alpha{}_\mu \gamma^B{}_{\nu\rho\sigma}
  - \triangleright_{\alpha B}{}^A  \alpha^\alpha{}_\nu \gamma^B{}_{\rho\sigma\mu}
  + \triangleright_{\alpha B}{}^A  \alpha^\alpha{}_\rho \gamma^B{}_{\sigma\mu\nu}
  - \triangleright_{\alpha B}{}^A  \alpha^\alpha{}_\sigma \gamma^B{}_{\mu\nu\rho} \vphantom{\ds\int} \\
 & & + 2X_{ab}{}^A \beta^a{}_{\mu\nu} \beta^b{}_{\rho\sigma}
  - 2X_{ab}{}^A \beta^a{}_{\mu\rho} \beta^b{}_{\nu\sigma}
   + 2X_{ab}{}^A \beta^a{}_{\mu\sigma} \beta^b{}_{\nu\rho}\,. \vphantom{\ds\int} \\
\end{array}
\end{equation}
Looking at the expressions for $\cG$ and $\cH$, one can note that it is always possible to combine the derivative term with the term containing a triangle into a covariant derivative term, as:
\begin{equation}
  \nabla_\lambda \beta^b{}_{\mu\nu} = \del_\lambda \beta^b{}_{\mu\nu} + \triangleright_{\alpha a}{}^b \alpha^\alpha{}_\lambda \beta^a{}_{\mu\nu} \,,\qquad
  \nabla_\mu \gamma^A{}_{\nu\rho\sigma} = \del_\mu \gamma^A{}_{\nu\rho\sigma} + \triangleright_{\alpha B}{}^A  \alpha^\alpha{}_\mu \gamma^B{}_{\nu\rho\sigma}\,,
\end{equation}
where the triangle combined with the connection 1-form $\alpha$ serves the purpose of the connection term for the covariant derivative. This suggests to introduce a notion of covariant exterior derivative (see equation (\ref{eq:jna8}) in the main text) as
\begin{equation} \label{eq:jna8druga}
\nabla = {\rm d} + \alpha \wedge^{\triangleright}
\end{equation}
which can act on any object in spaces $\Lambda_p(\cM) \otimes \mathfrak{g}$, $\Lambda_p(\cM) \otimes \mathfrak{h}$ and $\Lambda_p(\cM) \otimes \mathfrak{l}$. For example, given $\beta \in \Lambda_2(\cM) \otimes \mathfrak{h}$, we have:
\begin{equation}
\nabla \beta = {\rm d}\beta + \alpha \wedge^{\triangleright} \beta = \frac{1}{2} \Big[ \underbrace{\partial_\lambda \beta^b{}_{\mu\nu} + \triangleright_{\alpha a}{}^b \alpha^\alpha{}_\lambda \beta^a{}_{\mu\nu}}_{\nabla_\lambda \beta^b{}_{\mu\nu}} \Big] \; \rmd x^\lambda \wedge \rmd x^\mu \wedge \rmd x^\nu \otimes t_b \,,
\end{equation}
using the results of (\ref{eq:dbetaterm}) and (\ref{eq:alphabetaterm}). As another example, given $\phi \in \Lambda_0(\cM) \otimes \mathfrak{l}$ (i.e., a set of scalar fields $\phi = \phi^A T_A$), we have:
\begin{equation}
\nabla \phi = \rmd \phi + \alpha \wedge^\triangleright \phi = \partial_\lambda \phi^A \; \rmd x^\lambda \otimes T_A + \alpha^\alpha{}_\lambda \phi^B \; \rmd x^\lambda \otimes \tau_\alpha \triangleright T_B = \Big[ \underbrace{ \partial_\lambda \phi^A + \triangleright_{\alpha B}{}^A \alpha^\alpha{}_\lambda \phi^B}_{\nabla_\lambda \phi^A} \Big] \rmd x^\lambda \otimes T_A \,.
\end{equation}
The same calculation can be rewritten so that it does not expand 0-forms and 1-forms into a basis, as follows:
\begin{equation}
\nabla \phi = \rmd \phi + \alpha \wedge^\triangleright \phi = \rmd \phi^A \otimes T_A + \alpha^\alpha \wedge \phi^B \otimes \tau_\alpha \triangleright T_B = \Big[ \underbrace{ \rmd \phi^A + \triangleright_{\alpha B}{}^A \alpha^\alpha \wedge \phi^B}_{\nabla \phi^A} \Big] \otimes T_A \,.
\end{equation}
This illustrates equation (\ref{eq:jna9}) from the main text. As an exercise for an interested reader, one can apply (\ref{eq:jna8druga}) to rewrite the field strenghts (\ref{eq:ThreeCurvatureDefOpet}) into a more compact form (see equation (\ref{eq:ThreeCurvatureRewrite}) in the main text):
\begin{equation}
{\cal F}=\nabla^2 -\partial\beta\,,\qquad {\cal G}=\nabla\beta -\delta\gamma\,,\qquad
{\cal H}=\nabla\gamma +\{\beta\wedge\beta\}_{\rm pf}\,.
\end{equation}

Finally, let us rewrite the topological $3BF$ action into the component form. The action is defined as (see equation (\ref{eq:3BFaction}) in the main text):
\begin{equation}
S_{3BF}^\text{top}=\int_{\cM_4} \langle B\wedge {\cal F}\rangle_\mathfrak{g}+\langle C\wedge {\cal G}\rangle_\mathfrak{h}+\langle D\wedge {\cal H}\rangle_\mathfrak{l}.
\end{equation} 
The $G$-invariant nondegenerate symmetric bilinear forms $\langle\_\, ,\_\rangle_{\mathfrak{g}}$, $\langle\_\, ,\_\rangle_{\mathfrak{h}}$  and $\langle\_\, ,\_\rangle_{\mathfrak{l}}$  map a pair of algebra elements into a real number. Evaluating them on the basis vectors of the corresponding Lie algebras, one obtains their components:
\begin{equation}
\langle\tau_\alpha\, ,\tau_\beta\rangle_{\mathfrak{g}} = g_{\alpha\beta} \,, \qquad
\langle t_a\, , t_b \rangle_{\mathfrak{h}} = g_{ab}\,, \qquad
\langle T_A\, , T_B \rangle_{\mathfrak{l}} = g_{AB}\,.
\end{equation}
Keeping in mind that the Lagrange multipliers $B$, $C$ and $D$ belong to appropriate spaces,
\begin{equation}
B \in \Lambda_2(\cM)\otimes \mathfrak{g}\,, \qquad
C \in \Lambda_1(\cM)\otimes \mathfrak{h}\,, \qquad
D \in \Lambda_0(\cM)\otimes \mathfrak{l}\,,
\end{equation}
one can rewrite the action in terms of components as
\begin{equation}
S_{3BF}^\text{top}=\int_{\cM_4} \left( \frac{1}{4} g_{\alpha\beta} B^\alpha{}_{\mu\nu}\cF^\beta{}_{\rho\sigma} + \frac{1}{3!} g_{ab} C^a{}_\mu \cG^b{}_{\nu\rho\sigma} + \frac{1}{4!} g_{AB} D^A \cH^B{}_{\mu\nu\rho\sigma} \right) \rmd x^\mu \wedge \rmd x^\nu \wedge \rmd x^\rho \wedge \rmd x^\sigma \,.
\end{equation} 
Using the basic identity for differential forms
\begin{equation}
\rmd x^\mu \wedge \rmd x^\nu \wedge \rmd x^\rho \wedge \rmd x^\sigma = \lc^{\mu\nu\rho\sigma} \; d^4x\,,
\end{equation}
one can finally rewrite the action into the traditional form
\begin{equation}
S_{3BF}^\text{top}=\int_{\cM_4} \cL_{3BF}^\text{top} \;d^4x \,,
\end{equation} 
where the Lagrangian density for the topological $3BF$ theory is given in terms of component fields as:
\begin{equation}
\cL_{3BF}^\text{top}= \lc^{\mu\nu\rho\sigma} \left( \frac{1}{4} g_{\alpha\beta} B^\alpha{}_{\mu\nu}\cF^\beta{}_{\rho\sigma} + \frac{1}{3!} g_{ab} C^a{}_\mu \cG^b{}_{\nu\rho\sigma} + \frac{1}{4!} g_{AB} D^A \cH^B{}_{\mu\nu\rho\sigma} \right) \,.
\end{equation}

\end{document}